\documentstyle[psfig]{l-aa}
\newcommand{\cs}{\scriptstyle \rm }
\newcommand{\ha}{H$\alpha~$}
\newcommand{\hb}{H$\beta~$}
\newcommand{\hg}{H$\gamma~$}

\newcommand{\mpm}{$\pm$}

\newcommand{\msol}{M$_{\odot}$~}
\newcommand{\rsol}{R$_{\odot}$~}
\newcommand{\lsol}{L$_{\odot}$~}
\newcommand{\mdot}{\.{M}~}

\newcommand{\pyr}{yr$^{-1}$}
\newcommand{\mum}{$\mu$m~}
\newcommand{\vinf}{\mbox{$v_{\infty}$}~}

\begin{document}
\thesaurus{07%                   Stellar atmospheres
           (02.12.2;%            Line : identification
            08.01.3;%            Stars: atmospheres
            08.09.2: HD~101584;% Stars: individual
            08.13.2;%            Stars: mass loss
            08.16.1;%            Stars: peculiar
            08.16.4)}%           Stars: AGB and Post-AGB
\title{A study
on the nature of the peculiar supergiant HD~101584
\thanks{Based  on data collected at the European Southern
Observatory (La Silla, Chili), the International Ultraviolet Explorer,
IRAS and the Long-Term Photometry of Variables project.}
\thanks{Table~14 and 15 are also available in electronic form
at the CDS via anonymous ftp 130.79.128.5}}
\label{ch-art2}
\author{Eric J. Bakker\inst{1,2}\and
        Henny J.G.L.M. Lamers\inst{2,1}\and
        L.B.F.M. Waters\inst{3,4}\and
        Christoffel Waelkens\inst{5}\and
        Norman R. Trams\inst{6}\and
        Hans van Winckel\inst{5}}
\offprints{Eric J. Bakker, present address:
        Astronomy Department,
        University of Texas,
        Austin, TX 78712-1083,
        U.S.A., ebakker@astro.as.utexas.edu}
\institute{Astronomical Institute, University of Utrecht, P.O.Box 80.000,
           NL-3508 TA Utrecht,  The Netherlands\and
           SRON Laboratory for Space Research Utrecht,
           Sorbonnelaan 2, NL-3584 CA Utrecht, The Netherlands\and
           Astronomical Institute, University
           of Amsterdam, Kruislaan 403, NL-1098 SJ Amsterdam,
           The Netherlands\and
           SRON Laboratory for Space Research Groningen, P.O.Box 800,
           NL-9700 AV Groningen, The Netherlands\and
           Astronomical Institute, Katholieke Universiteit Leuven,
           Celestijnenlaan 200, B-3030 Heverlee, Belgium\and
           ESTEC Space Science Department, P.O.Box 299, NL-2200 AG
           Noordwijk, The Netherlands}
\date{received May 24 1995, July 15 1995}
\maketitle
\begin{abstract}

 We present a study of low- and high-resolution ultraviolet,
 high-resolution optical CAT/CES spectra  and ultraviolet, optical
 and infrared photometry of the peculiar
 supergiant HD~101584.
 From the photometry we learn that the ultraviolet and optical energy
 distribution cannot be fitted in
 a consistent way and we need a model
 in which the UV and optical energy distribution are formed by different
 gas.  The Geneva photometry is best fitted to a B9II Kurucz model,
 $T_{eff}=12000\pm1000$~K and $\log g=3.0\pm1.0$, with an extinction of
 $E(B-V)=0.49\pm0.05$.

 The observed spectral features in the spectrum of HD~101584 are
 classified in eight different categories based on the
 velocity, shape of profile and the identification.
 The high-excitation
 HeI ($\chi=20.87$~eV), NII ($\chi=18.40$~eV), CII ($\chi=14.39$~eV)
 and NI ($\chi=10.29$~eV) optical absorption lines are formed
 in the photosphere of a late B-star (e.g. B8-9I-II).
 These absorption lines show radial velocity variations which are
 attributed to binary motion, with the secondary being a white dwarf
 or a low-mass main sequence object.
 The low-excitation
 P-Cygni lines in the optical and UV are formed in the wind. The number
 density of absorption lines in the UV is so large that the
 wind spectrum acts as an iron curtain in front of the B-star. The
 terminal velocity of the wind of $v_{\infty}=100\pm30$~km~s$^{-1}~$  is
 consistent with the star being a low-mass post-AGB star and the low
effective gravity is attributed to the presence of a nearby,
unseen, secondary. We estimate a mass-loss rate
 of \.{M}$\approx 10^{-8}$~M$_{\odot}$~yr$^{-1}$.
 Narrow absorption and emission lines are observed which are formed
 in a circumsystem disk with a typical radius of $10^{2} R_{\ast}$.

\end{abstract}

\section{Introduction}

 We have conducted a multiwavelength study of the enigmatic object HD~101584.
 Classified as a F0Iape (Hoffleit~{\sl et al.}
 \cite{art2hoffleitetal}) it is not a
 normal F0-type supergiant. The spectrum has many peculiarities, ranging from
 very narrow  optical emission lines of neutral
 and singly ionized metals (Trams~{\sl et al.} \cite{art2tramsetal1}), P-Cygni
 profiles,
 the HeI 5876~\AA~ absorption line  (Morrison and Zimba \cite{art2morrison})
 to a very large IR excess due to dust
(Humphreys and Ney \cite{art2humphreysney};
 Parthasarathy and Pottasch \cite{art2parthasarathypottasch};
Oudmaijer~{\sl et al.} \cite{art2oudmaijer}).
 At millimeter wavelengths we find molecular line
 emission of CO
 (Trams~{\sl et al.} \cite{art2tramsetal1};
 Loup~{\sl et al.} \cite{art2loupetal};
 Van der Veen~{\sl et al.} \cite{art2veenetal})
 and OH
 (Te Lintel Hekkert~{\sl et al.} \cite{art2lintelhekkertetal}).
 The very complex structure of the CO emission
 shows large Doppler velocities of
 130~km~s$^{-1}$ with respect to the central velocity of the feature,
 indicating
 outflow velocities of more than 100~km~s$^{-1}$.
 The OH maser line (1667~MHz) emission shows much lower velocities of
 30~km~s$^{-1}$ with respect to the same central velocity of
 50\mpm2~km~s$^{-1}$.

 In the last two decades two models have been developed for
 HD~101584 and in both models binarity is suggested to explain the
 observations.

 The first serious model was
 introduced by Humphreys and Ney (\cite{art2humphreysney}). They argue that
 the infrared radiation is due to the secondary in a binary system.
 This secondary fills its Roche lobe and material is transferred through
 the inner Lagrangian point and accreted by the primary. We will show in
 this paper that this model cannot be supported because we find overwhelming
 evidence for outflow, rather than for in fall.

 A second model was suggested by Parthasarathy and Pottasch
 (\cite{art2parthasarathypottasch}) and supported by
 Trams~{\sl et al.}  (\cite{art2tramsetal2}). In this model HD~101584 is in the
 evolutionary stage of the Post-Asymptotic Giant Branch (post-AGB). During
 the preceding phase of its evolution on the AGB the star had a
 strong stellar
 wind. After the star has left  the AGB the expelled material
 (the AGB remnant) slowly moves away from the star.
 The dust is in radiative equilibrium with
 the radiation field of the star, and re-radiates the absorbed energy
 in the infrared part of the spectrum.
 The post-AGB nature of HD~101584 is supported by
 the space velocity of
 the star derived from the central velocity of the CO and OH millimeter
 line emission. This velocity of $v_{rad}=50.3\pm2.0$~km~s$^{-1}$
 does not agree with the galactic rotation curve
 if the star is at the distance derived by assuming that it is a massive
 Population I star.

 In an attempt to solve the many controversies about this enigmatic star,
 we have made a multiwavelength study of the ultraviolet, optical and
 infrared part of the spectrum.
 Based on these observations we propose a new model for HD~101584.

 Sect.~\ref{art2sec-obs} gives information about the spectral
and photometric data
 collected and used in this study. In Sect.~\ref{art2sec-energy} we look
 at the total energy distribution and try to fit it with a F0I Kurucz model.
 In Sect.~\ref{art2sec-cat} we  distinguish eight different
 categories of spectral lines and derive
 for each category a characteristic velocity. In Sect.~\ref{art2sec-nature}
 we converge
 to consistency between the results of Sect.~\ref{art2sec-energy}
 and Sect.~\ref{art2sec-cat} and discuss the nature of HD~101584.
 Sect.~\ref{art2sec-conc} gives the conclusions.

\section{The observations}
\label{art2sec-obs}
 We have collected ultraviolet (IUE), optical (ESO) and infrared (IRAS)
 spectra completed with
 optical (Geneva, Str\"{o}mgren and Johnson) and infrared
 (near-infrared and IRAS)  photometry.

 \subsection{Ultraviolet spectra}

 The low-resolution (LORES) ultraviolet
 spectra of HD~101584 have a wavelength coverage
 from 1190~\AA~ to 3200~\AA~ with a resolution of
 $\Delta \lambda = 6.1$~\AA~ for the
 short-wavelength spectra, and $\Delta \lambda =9.1$~\AA~
 for the long-wavelength
 spectra. The LORES spectra of HD~101584 were obtained from the IUE archive
 (SWP31025, LWP10808 and LWP10809 Table~\ref{art2tab-uvspectra}).
 A description of the IUE satellite can be found
 in  Boggess~{\sl et al.} (\cite{art2boggessa} and \cite{art2boggessb})
 The absolute flux calibration
 of LORES spectra makes it possible to combine it with optical
 and infrared photometry
 to make an energy distribution from 1190~\AA~ to 100~$\mu$m and
 to compare the shape of the spectral energy distribution (SED)
to Kurucz model atmospheres.

 \begin{table}
 \caption{Ultraviolet spectra of HD101584}
 \label{art2tab-uvspectra}
 \centerline{\begin{tabular}{lllll}
 \hline
 Camera    &Observation&\multicolumn{3}{c}{Wavelength range}  \\
 Image No. &Date       &\multicolumn{3}{c}{[\AA]}             \\
 \hline
 \multicolumn{5}{c}{Low-resolution IUE spectra} \\
 \hline
          &              &         & &         \\
 SWP31025 & 22 May  1987 & 1191    &-& 1950    \\
 LWP10808 & 22 May  1987 & 1951    &-& 2551    \\
 LWP10809 & 22 May  1987 & 2552    &-& 3199    \\
          &              &         & &         \\
 \hline
 \multicolumn{5}{c}{High-resolution IUE spectra}\\
 \hline
          &              &         & &      \\
 LWP17369 & 15 Febr 1990 & 2500    &-& 3000 \\
 LWR04822 & 21 June 1979 & 2500    &-& 3000 \\
          &              &         & &      \\
 \hline
 \end{tabular}}
 \end{table}

 We used two high-resolution (HIRES) IUE spectra in the long-wavelength
 range with $\Delta \lambda$=0.3~\AA.
 One HIRES spectrum LWP~17369 was observed by one of
 us (NRT), whereas the other, LWR04822, was retrieved
 from the archive at the IUE ground station VILSPA (Spain)
 (Table~\ref{art2tab-uvspectra}).

 For the identification of the high-resolution UV spectra we made
 use of a high-resolution spectrum of $\alpha$~Lep. This F0I
 type star has the  same spectral classification as HD~101584.
 Comparing the two spectra will enable us to make  a quantitative comparison
 (Sect.~\ref{art2sec-IV}).

 \subsection{Optical spectra}

 The optical spectra were obtained with the 1.4 meter
 Coud\'{e} Auxiliary Telescope (CAT)
 at ESO using the Coud\'{e} Echelle Spectrograph (CES)
 by CW and HvW. Each
 spectrum has a coverage of  about 50~\AA~ with a resolution of
 $\Delta\lambda =0.1$~\AA. These spectra have no absolute flux calibration,
 so they were normalized to the continuum.
 Table~\ref{art2tab-spectra} gives a list of the optical spectra used in this
 study in order of increasing wavelength.

 \begin{table*}
 \caption{CAT/CES spectra of HD~101584}
 \label{art2tab-spectra}
 \centerline{\begin{tabular}{lllrllll}
 \hline
 \multicolumn{3}{c}{Observation}&$\delta v_{\oplus}^{*}$&
 \multicolumn{3}{c}{Wavelength range}& Remark  \\
 Date         & U.T.& J.D.      &[km~s$^{-1}$]    &
 \multicolumn{3}{c}{[\AA]}        &     \\
 \hline
               &     &         &      &    & &    &               \\
  20 Jan. 1990 &8h20 & 7911.847&+19.08&4284&-&4318&               \\
  20 Feb. 1989 &8h06 & 7577.838&+15.50&4327&-&4357& \hg           \\
  21 Jan. 1990 &7h43 & 7912.822&+19.09&4373&-&4407&               \\
  20 Jan. 1990 &8h41 & 7911.861&+19.06&4463&-&4498& MgII          \\
  20 Feb. 1989 &7h42 & 7577.821&+15.53&4842&-&4877& \hb           \\
  19 Feb. 1989 &6h28 & 7576.847&+15.68&5025&-&5064& NII \& HeI    \\
  13 Feb. 1993 &5h55 & 9031.747&+16.91&5859&-&5911& NaI D \& HeI  \\
  12 Feb. 1993 &6h33 & 9030.770&+17.02&6537&-&6593& CII \& \ha    \\
  20 Apr. 1992 &5h34 & 8732.732&-01.25&7413&-&7476& NI            \\
  13 Feb. 1993 &4h12 & 9031.675&+17.01&8649&-&8721& NI \& CaII    \\
               &     &         &      &    & &    &               \\
 \hline
 \end{tabular}}
 \begin{centering}
 $^{*}$ The correction term to convert from observed
 $v_{obs}$ to heliocentric radial velocity  $v_{\odot}$:
 \( v_{\odot} = v_{obs} + \delta v_{\oplus}\) \newline
 To convert to local standard of rest (LSR) velocities
 \( \delta v_{\odot} = 9.22 \)~km~s$^{-1}$
 should be added to $v_{\odot}$ \newline
 \end{centering}
 \end{table*}

 \subsection{Infrared spectra}

 The low-resolution spectrum (LRS) observed with  IRAS was obtained
 from the archive at Space Research Groningen. The flux scale
 is absolutely calibrated and the wavelength  coverage is
 from 7 to 24~$\mu$m.

 \subsection{Photometry}

 From a large set of photometric data available for HD 101584
we have selected the  data-points
 for which the observation date is closest to the observation date
 of the LORES UV spectra (Table~\ref{art2tab-photo}).
 In this way we minimize the effect of photometric
 variability of HD~101584 on the composed energy distribution.
 The UV flux is calculated from the IUE spectra by binning the
 spectrum in 10 wavelength intervals. The bin with the lowest central
 wavelength has been rejected as this bin only contains noise.
 The Geneva photometry is collected by one
 of us, CW, whereas the Str\"{o}mgren photometry is taken from the Long
 Term Photometric of Variables (LTPV) project
 (Manfroid~{\sl et al.} \cite{art2manfroidetal};
 Sterken~{\sl et al.} \cite{art2sterkenetal}).
 The Johnson UVBRI and near-infrared JHKLM
 photometry are from Trams~{\sl et al.} (\cite{art2tramsetal2}),
 and the IRAS fluxes from the IRAS
 point source catalogue (JISWG \cite{art2iras}).
 The observed magnitudes are converted
 to a flux scale using the calibration  of
 Rufener and Nicolet (\cite{art2rufenernicolet}) for the Geneva filters,
 Manfroid~{\sl et al.}
(\cite{art2manfroidetal}) for the LTPV Str\"{o}mgren filters,
 Landolt-B\"{o}rnstein (page 73, \cite{art2landoltbornstein})
 for visual Johnson-photometry
 and  Korneef (\cite{art2korneef}) for near-infrared photometry.
 The IRAS fluxes are color corrected as described in the
 IRAS Explanatory Supplement (JISWG \cite{art2iras}).

 \begin{table}
 \caption{Observed photometry of HD~101584}
 \label{art2tab-photo}
\begin{small}
 \centerline{\begin{tabular}{lllll}
 \hline
 Date         &$\lambda_{c}$&  &Magn. & log flux$^{\ast}$ \\
 \hline
 \multicolumn{5}{c}{Ultraviolet photometry} \\
 \hline
             &          &     &      &        \\
 22 May 1987 &1365~\AA  &     &      & -26.06 \\
 JD=6938     &1493      &     &      & -25.28 \\
             &1640      &     &      & -24.84 \\
             &1786      &     &      & -24.42 \\
             &1939      &     &      & -24.25 \\
             &2154      &     &      & -24.25 \\
             &2387      &     &      & -24.17 \\
             &2619      &     &      & -23.75 \\
             &2851      &     &      & -23.45 \\
             &3082      &     &      & -23.09 \\
 \hline
 \multicolumn{5}{c}{Geneva photometry} \\
 \hline
 29 Feb. 1987&3450~\AA  &$U $& 7.841& -22.78 \\
 JD=6865.731 &4020      &$B1$&  7.322& -22.37 \\
             &4250      &$B $& 6.421& -22.33 \\
             &4480      &$B2$& 7.886& -22.29 \\
             &5400      &$V1$& 7.637& -22.19 \\
             &5500      &$V $& 6.913& -22.19 \\
             &5810      &$G $& 8.015& -22.18 \\
 \hline
 \multicolumn{5}{c}{LTPV Str\"{o}mgren photometry} \\
 \hline
             &3500~\AA  &$u$ & 8.554& -22.34 \\
 5 July 1986 &4110      &$v$ & 7.517& -22.09 \\
 JD=6617.514 &4670      &$b$ & 7.235& -22.11 \\
             &5470      &$y$ & 6.905& -22.19 \\
 \hline
 \multicolumn{5}{c}{Visual Johnson UVBRI band photometry} \\
 \hline
 30 Jan. 1987&3600~\AA  &$U_j$& 7.308& -22.67 \\
 JD=6836     &4400      &$B_j$& 7.372& -22.30 \\
             &5500      &$V_j$& 6.955& -22.22 \\
             &7000      &$R_j$& 6.689& -22.23 \\
             &9000      &$I_j$& 6.389& -22.21 \\
 \hline
 \multicolumn{5}{c}{Near-infrared photometry} \\
 \hline
 2 Feb. 1987 &1.25~$\mu$m&$J$ & 5.975& -22.18 \\
 JD=6839     &1.65      &$H$  & 5.152& -22.02 \\
             &2.20      &$K$  & 3.831& -22.71 \\
             &3.80      &$L$  & 1.791& -21.27 \\
             &4.70      &$M$  & 1.112& -21.23 \\
 \hline
 \multicolumn{5}{c}{IRAS photometry} \\
 \hline
             &  12~$\mu$m&    &88.7~Jy& -21.05 \\
             &  25      &     &133.7 & -20.88 \\
             &  60      &     &177.4 & -20.75 \\
             & 100      &     & 98.6 & -21.01 \\
             &          &     &      &        \\
 \hline
 \end{tabular}}
 \centerline{$^{\ast}$ in $\rm erg~cm^{-2}~s^{-1}~Hz^{-1}$}
\end{small}
 \end{table}

 \section{Spectral type from photometry}

 \label{art2sec-energy}
 \subsection{The total energy distribution}

\begin{figure}
\centerline{\hbox{\psfig{figure=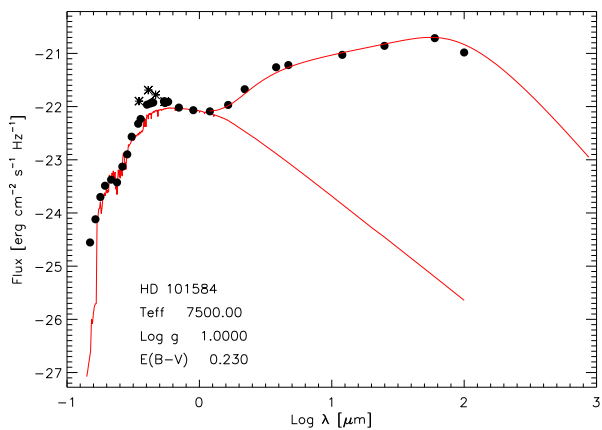,width=\columnwidth}}}
\caption{The complete dereddened
         energy distribution of HD~101584 with $E(B-V)=0.230$, ranging
         from the ultraviolet (1190~\AA) to the far-infrared
         (100~$\mu$m). The solid line is a $T_{\rm eff}=7500$~K,
         $\log g=1.0$
         Kurucz model. Note the excess of energy at
         the Balmer jump and at the infrared. The Str\"{o}mgren
         photometry is plotted with an asterisk. Their higher
         flux is due to photometric variability of the star}
\label{art2fig-sed}
\end{figure}

 The energy distribution of HD~101584 is fitted to a
 Kurucz model atmosphere (Kurucz \cite{art2kurucz}) with parameters
 $T_{eff}=7500$~K and $\log g=1.0$
 which are appropriate for the spectral type
 of F0I (Hoffleit~{\sl et al.} \cite{art2hoffleitetal}) and
 the corresponding reddening of $E(B-V)=0.23$. Fig.~\ref{art2fig-sed}
 shows that we are not able to fit the ultraviolet and optical
 energy distribution in a consistent way. After normalization at
 the I-band there is an excess of flux shortward of 1500~\AA~
 ($\log \lambda \leq -0.8$ in $\mu$m)  and
 an excess at the Balmer  jump at 3653~\AA~ ($\log \lambda = -0.4$).
 The excess shortward of 1500~\AA~
 indicates that the star is hotter than F0I.

The huge IR excess indicates the presence of dust.
We have tried to fit
an optically thin spherically symmetric dust model
to the complete energy distribution
(Waters~{\sl et al.} \cite{art2watersetal88})
(see Fig.~\ref{art2fig-sed}).
{}From this fit we find that the dust is at  a distance between
20  and $10^{5} R_{\ast}$ with dust temperatures between
$T_{\rm dust}=50$ and 1240~K. If HD~101584 is a 0.54~\msol
post-AGB star  (R$_{\ast}=38$~R$_{\odot}$ and $\log L = 3.6$,
Boothroyd and Sackmann, \cite{art2bootsack}) than this
yields a total dust mass of $M_{\rm dust} = 2\times10^{-2}$~\msol.
Taking a gas-to-dust ratio of a hundred
gives a total circumstellar mass of $M_{CS}=2$~\msol.

The total amount of observed energy radiated in the infrared of
$E= 9.4 \times 10^{-8}$~erg~cm$^{-2}$s$^{-1}$ can be provided by
a reddening of $E(B-V)=0.39\pm0.02$ from a spherically symmetric
dust shell. This means that the central star has a $(B-V)_{o} \leq 0.02$
and a spectral type earlier than A0.

We conclude that the optical and UV energy distribution of HD~101584
indicates that the star (or one of the components if it is a binary)
is hotter than $T_{eff} \simeq 7500$~K, which is the temperature
suggested by its spectral type of F0I. We will show below that the
Geneva photometry and the
presence of high-excitation lines in the optical spectrum support this.

\subsection{The spectral type from the Geneva photometry}

The optical photometry can be used to derive the spectral type of the star.
Using the
Geneva photometry (Table~\ref{art2tab-photo})
and the calibration of the Geneva extinction-free parameters
$X$, $Y$ and $Z$  by Cramer and Maeder (\cite{art2cramermaeder}) we
are able to determine the $T_{eff}$ and $\log g$ of the star.
The values for HD~101584 are $X=1.150$, $Y=0.140$ and $Z=0.0219$.
The value of the $Z$ parameter can be used as a test to determine
peculiarities, because the
$Z$ parameter is close to zero for normal star while it
deviates  from zero for peculiar stars e.g. HgMn, Bp and Ap stars.
The low value of $Z$ for HD~101584 suggests that the
star is not peculiar in the $XYZ$ space. This encourages us to
use predicted colors for normal atmospheres to derive the stellar
parameters from the Geneva photometry.

\begin{figure}
\centerline{\hbox{\psfig{figure=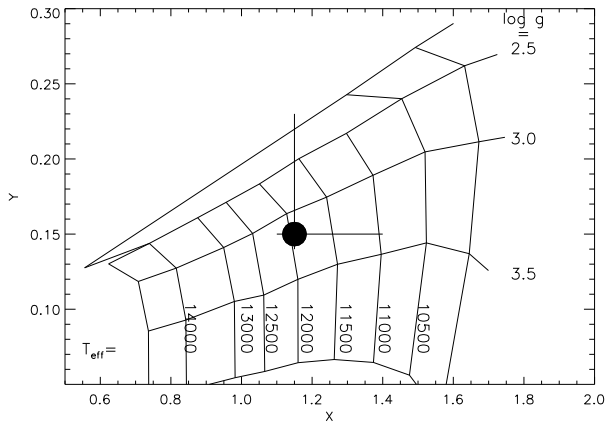,width=\columnwidth}}}
\caption{The Geneva extinction free $XY$-plane with the position of
         HD~101584. The grid has been computed with solar metalicity
         Kurucz models. The best fit is reached for
         $T_{eff}=12000\pm1000$~K, $\log g=3.0\pm1.0$
         and $E(B-V)=0.49\pm0.05$. Note that the error is largest
         in the direction of lower $\log g$ and lower $T_{eff}$}
\label{art2fig-xyz}
\end{figure}

We have computed the location of atmospheric models with solar
metalicity (Kurucz \cite{art2kurucz}) in the
the $XY$-diagram. (Fig.~\ref{art2fig-xyz}). The location of HD~101584
is closest to the Kurucz model with
$T_{eff}=12000$~K, $\log g=3.0$ and $E(B-V)=0.49$.

We will make a rough estimate of the possible error on the derived
$X$ and $Y$ of the star
caused by the presence of wind lines in the Geneva bands.
Only the $U$ and  $V1$ band
have enough wind lines that the photometry might be effected
(the $V$ band is not used for calculating the $X$, $Y$ and $Z$
parameters).
A strong blocking by wind lines would increase the magnitudes
of the star compared to normal stars of the same type. A decrease in flux
of 5\% in only the $U$ band
gives a lower temperature of $T_{eff}=11500$~K and a
surface gravity of $\log g=2.5$, whereas a decrease in flux
of 5\% in only the $V1$
magnitude yields $T_{eff}=11000$~K and $\log g \leq 2.0$. A decrease of 5\%
in both bands gives $T_{eff}=10000$~K and $\log g \leq 2.0$.
This shows that the presence of
wind lines in the optical spectrum changes the best fit in the $XY$-diagram
only slightly. We adopt
$T_{eff}=12000\pm1000$~K, $\log g=3.0\pm1.0$,
and $E(B-V)=0.49\pm0.05$  as the best fit to the Geneva
$X$ and $Y$ parameters. This corresponds to a B9II star.

We will show below that this temperature is in good agreement
 with the spectral type found from photospheric optical
absorption lines. The effective gravity which is most
affected by blanketing by wind lines  is rather
uncertain.

\section{Line-identification, line profiles and velocities}
\label{art2sec-cat}

 In this section we will look at the ultraviolet, optical and infrared
 spectra to learn about different categories of lines which
 are seen in these spectra.  The high-resolution IUE spectrum has
 been studied by Bakker (\cite{art2bakkerart1}) and a selected list of
 line identifications which will be used in this work
 is given in Table~\ref{art2tab-fe} of App.~\ref{art2ap-uvFeII}.
 We have made a line identification of the optical spectra
 in  selected wavelength regions (Table~\ref{art2tab-spectra}).
 The identification of the optical spectra
 as given the Table~\ref{art2tab-optical} (App.~\ref{art2ap-optical})
 shows absorption lines from
 high-excitation ($\chi \geq 10.0$~eV),
 low-excitation  ($\chi \leq 10.0$~eV) levels, emission lines from
 neutral- and singly ionized metals and complex hydrogen line profiles.

 The observed spectral features are classified in eight different
 categories on the basis
 of the excitation conditions of the atom, shape and velocity of the
 line profile. To allow comparison between different categories, and
 observation from different dates, all velocities quoted are heliocentric and
 to be able to determine velocities relative to the system (star), we have
 adopted the central velocity of the CO and OH emission, i.e.
 +50\mpm2~km~s$^{-1}$,
 as the system velocity (see also Sect.~\ref{art2sec-VIII}).

\begin{figure*}
\centerline{\hbox{\psfig{figure=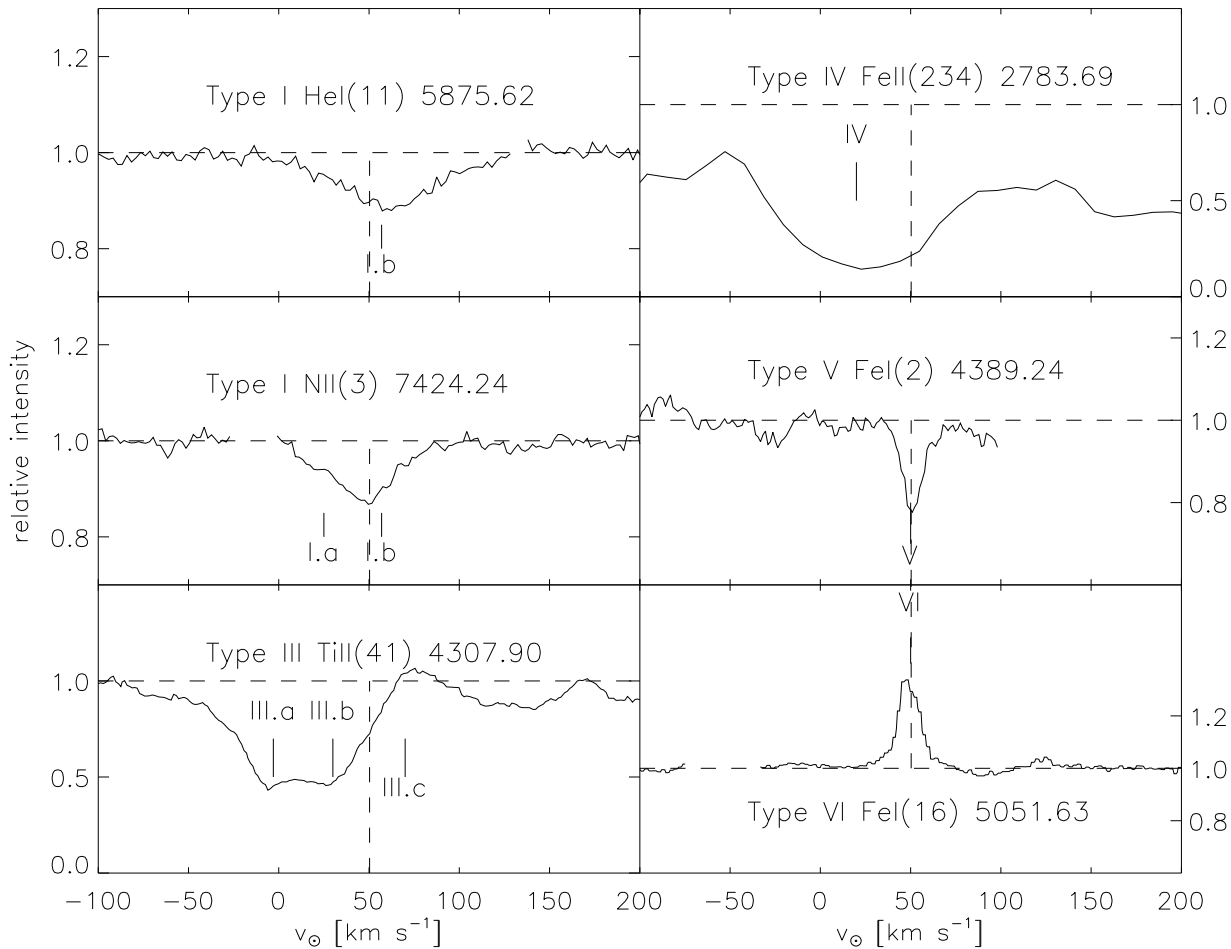,width=\textwidth}}}
\caption{Examples of line types  I, III, IV, V and VI}
\label{art2fig-lines1}
\end{figure*}

 We distinguish the following types of spectral lines
 (Table~\ref{art2tab-linetype}), where we have
 ordered the categories of lines according to their excitation conditions.
 Starting with high-excitation lines going down to lower-excitation levels and
 finishing with molecular lines. In the next section we will argue that
 this sequence of lines is also a sequence in geometric distance between
 the line forming region and the central object.
 Fig.~\ref{art2fig-lines1} and
 \ref{art2fig-lines2} shows examples
 of the different categories of lines.

\begin{table*}
\caption{Eight different type of spectral lines are distinquished}
\label{art2tab-linetype}
\centerline{\hbox{\begin{tabular}{ll}
\hline
Type             & Description                              \\
\hline
                 &                                          \\
I    & optical pure absorption lines from high-excitation
       levels                                                \\
II   & Balmer lines                                          \\
III  & optical P-Cygni lines from low-excitation  levels of
       neutral- and singly ionized metals                    \\
IV   & ultraviolet absorption lines from low-excitation
       levels of neutral- and singly ionized metals          \\
V    & optical pure absorption lines from low-excitation
       levels of neutral  metals                             \\
VI   & optical pure emission lines from low-excitation
       levels of neutral and singly ionized metals           \\
VII  & infrared emission feature (10~\mum)                   \\
VIII & molecular emission lines (CO and OH)                  \\
     &                                                       \\
\hline
\end{tabular}}}
\end{table*}

\subsection{Type I: Optical pure absorption lines
                    from high-excitation levels}
\label{art2sec-I}

We have detected absorption lines of CII (14.39~eV),
HeI (20.87~eV), NII (18.40~eV) and NI (10.29~eV), in the optical
spectrum of HD~101584.  This is not consistent with the spectral
type of F0Iape  as given by Hoffleit~{\sl et al.} (\cite{art2hoffleitetal}).

The presence of the CII lines suggests that the optical
spectrum (or at least part of it) is produced by a B-type star.
For supergiants the CII lines are only formed between
types B0.5I and B9I, and reach a
maximum strength in B3Ia-type stars (Lennon~{\sl et al.} \cite{art2lennonetal1}
and \cite{art2lennonetal2}).
The  observed equivalent widths measured for CII(6578) and CII(6583)
of  85 and 45~m\AA~ respectively are of the same order as the equivalent width
of a B8.5I and B0.5I star. However as we have also detected NI at
the same velocity as HeI and CII
we can eliminate the possibility of an early-B type star.

We can obtain additional information by studying the radial velocity
of the high-excitation lines.
Table~\ref{art2tab-velI} gives the observation dates, the average radial
velocity on that date and the number of spectral lines (type I)
observed on that
date. The average radial velocity of all observation dates is
56.9\mpm5.9~km~s$^{-1}$.  We notice that the
difference in velocity between successive observations is much larger
than the error ($\approx 7 \sigma$) in the individual measurements.
As it is unlikely that these high-excitation lines are formed
in the stellar wind it is suggestive to attribute
the radial velocity variations to the
orbital motions of the late-B star around an object whose nature
still has to be determined. The amplitude of the
radial velocity variations around the
system velocity of 50~km~s$^{-1}$
 is at least 10~km~s$^{-1}$. The data for the radial
velocity measurements are too scarce to estimate the orbital period.

 \begin{table}
 \caption{Radial velocities of type I lines}
 \label{art2tab-velI}
 \centerline{\begin{tabular}{lllll}
 \hline
           & \multicolumn{2}{c}{abs. I.a}
           & \multicolumn{2}{c}{abs. I.b}         \\
 Date      & $v_{\odot}$  &$N$ & $v_{\odot}$  &$N$ \\
 \hline
           &              &    &              &    \\
 Feb. 1989 &              & 00 & 44.0\mpm2.1  & 02 \\
 Jan. 1990 &              & 00 &              & 00 \\
 Apr. 1992 & 24.7\mpm1.7  & 03 & 54.0\mpm1.7  & 03 \\
 Feb. 1993 &              & 00 & 60.8\mpm0.6  & 09 \\
           &              &    &              &    \\
Average    & 24.7\mpm1.7  & 03 & 56.9\mpm5.9  & 14 \\
           &              &    &              &    \\
\hline
\end{tabular}}
\end{table}

Of all the high-excitation absorption lines observed only the
three NI(mp=3) lines in April 1992 show a second absorption component
at 24.7\mpm1.7~km~s$^{-1}$ (Table~\ref{art2tab-velI}).
This suggests that the high-excitation absorption
lines not only have a variable Doppler velocity, but also a variable
line profile.

We conclude that the high-excitation lines in the
optical spectrum of HD~101584 are produced
in the photosphere of a late-B type star.
Its $T_{eff}$ would be between about 10~500~K (B9) and 14~000~K (B5).
This is in agreement with the earlier conclusion that the
energy distribution indicates $T_{eff} \geq 7500$~K and with the spectral
type of B8-9 derived from the Geneva photometry.
The star shows radial velocity variations of at least 10~km~s$^{-1}$, which
probably indicates the presence of a companion.

\subsection{Type II: Hydrogen line profiles and the terminal
                     outflow velocity}

\label{art2sec-II}

\begin{figure*}
\centerline{\hbox{\psfig{figure=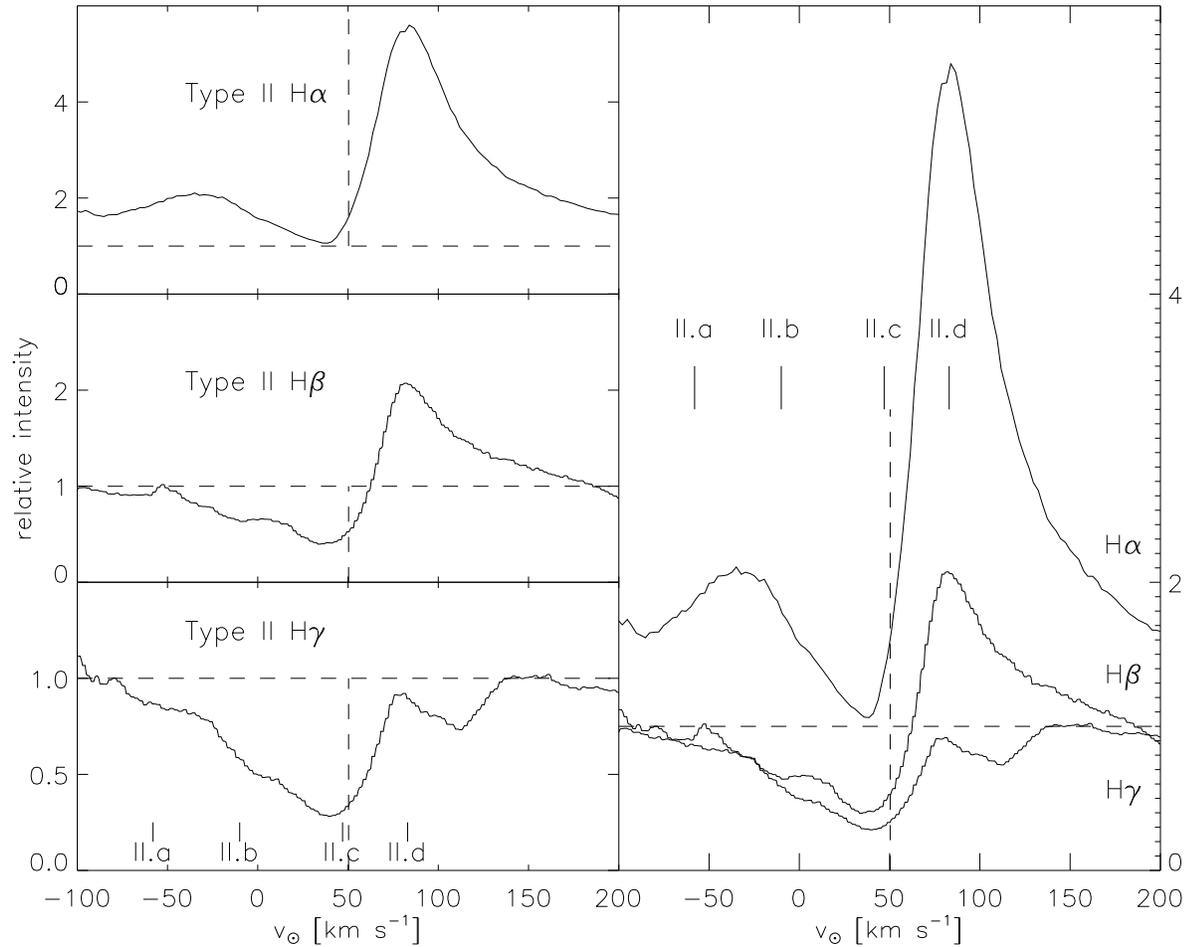,width=\textwidth}}}
\caption{Hydrogen \ha, \hb and \hg line profiles. Notice the second
         absorption component at 0~km~s$^{-1}$}
\label{art2fig-lines2}
\end{figure*}

The Balmer lines have a P-Cygni type profile (see Fig.~\ref{art2fig-lines2}
and Fig.~\ref{art2fig-opt1}, \ref{art2fig-opt2} and
\ref{art2fig-opt3})  with a strong emission component (\ha and \hb)
and a blue-shifted absorption component (indicated as II.c in
Fig.~\ref{art2fig-lines2}). These lines are formed in the stellar wind.

\begin{table*}
\caption{Radial velocities of type II lines}
\label{art2tab-velII}
\centerline{\begin{tabular}{llllll}
\hline
          &   &abs. II.a  &abs. II.b  &abs. II.c  &em. II.d    \\
Date      &   &$v_{\odot}$&$v_{\odot}$&$v_{\odot}$&$v_{\odot}$ \\
\hline
          &               &           &           &            \\
Feb. 1989 &\hg&-58        & -10       & 44        & 78         \\
          &\hb&-58        & -18       & 60        & 82         \\
Jan. 1990 &   &           &           &           &            \\
Apr. 1992 &   &           &           &           &            \\
Feb. 1993 &\ha&           &  -1       & 37        & 83         \\
          &   &           &           &           &            \\
Average   &   &-58\mpm2.1 &-10\mpm4.9 &47\mpm4.9  &81\mpm1.5   \\
          &   &           &           &           &            \\
\hline
\end{tabular}}
\end{table*}

The profile of \ha is complex: a P-Cygni profile superimposed
on a very broad emission profile. The emission extends from
about -1000 to + 1000~km~s$^{-1}$
(see also Fig.~\ref{art2fig-opt2} where the width of the
\ha emission is
of the same order as the total wavelength coverage of the spectrum).
It is most likely due to electron scattering by free electrons in the
stellar wind. The absorption wing
of \ha extends to heliocentric velocity of about -30~km~s$^{-1}$,
which corresponds to
-80~km~s$^{-1}$ in the frame of the system. The emission component of \hb
extends to about +180~km~s$^{-1}$, i.e. +130~km~s$^{-1}$
in the frame of the system.
This suggests that the P-Cygni profiles are formed in a wind which have a
maximum outflow velocity of about 100\mpm30~km~s$^{-1}$.
We will show below that
this is similar to the terminal velocity of about 100~km~s$^{-1}$ derived from
the P-Cygni profiles of lines from ionized metal in the UV and in the visual,
and is similar to the CO outflow velocity.

The profiles of
\hb and \hg show evidence for two additional absorption components
(indicated as II.a and II.b in Fig.~\ref{art2fig-lines2} and
Table~\ref{art2tab-velII}) at heliocentric velocities of
-58\mpm2 and -10\mpm5~km~s$^{-1}$, i.e. at -108 and -60~km~s$^{-1}$
in the frame of the system.
These components resemble the discrete absorption components that are
commonly observed in the profiles of lines formed in stellar winds
(Henrichs \cite{art2henrichs})
and they are interpreted as the result of variable mass loss or
density concentrations in the wind.

\subsection{Type III: Optical P-Cygni lines
                      from low-excitation levels of neutral
                      and singly ionized metals}
\label{art2sec-III}

In this study we define a line as a P-Cygni type  line if the spectral
feature shows an absorption (III.b) and an emission (III.c) component with the
emission on the long wavelength side of the absorption.
The spectrum of HD~101584 shows P-Cygni profiles of lines from
low-excitation levels of neutral and singly ionized metals,
e.g. FeI, TiI, MnI, NiI and TiII.

Most of the P-Cygni lines show a narrow emission peak (Type VI) superposed
on the wider emission of the P-Cygni profile.
In some
cases is was not possible to distinguish between the pure emission and
the emission
of the P-Cygni profile. In those cases the two emissions blend together.
The weakest P-Cygni type lines clearly show the presence of
a second, blue-shifted absorption component (III.a)
while for the stronger lines the two
absorption components blend together to one broad profile
(Fig.~\ref{art2fig-lines1}).

 \begin{table*}
 \caption{Radial velocities of type III lines}
 \label{art2tab-velIII}
 \centerline{\begin{tabular}{lllllll}
 \hline
          &\multicolumn{2}{c}{abs. III.a}
          &\multicolumn{2}{c}{abs. III.b}
          &\multicolumn{2}{c}{em.  III.c} \\
 Date     &$v_{\odot}$   &$N$&$v_{\odot}$   &$N$&$v_{\odot}$   &$N$\\
 \hline
          &              &  &               &  &               &  \\
 Feb. 1989 & +0.4\mpm3.2  &09& 31.8\mpm2.4   &17& 62.3\mpm4.6   &13\\
 Jan. 1990 & -4.3\mpm1.0  &27& 30.3\mpm1.1   &37& 72.9\mpm4.3   &21\\
 Apr  1992 &              &00&               &00&               &00\\
 Feb. 1993 & -3.5\mpm1.4  &04& 24.0\mpm4.8   &05& 76.3\mpm2.3   &04\\
           &              &  &               &  &               &  \\
 Average   & -3.1\mpm1.8  &40& 30.2\mpm2.1   &59& 69.6\mpm4.2   &38\\
           &              &  &               &  &               &  \\
 \hline
 \end{tabular}}
 \end{table*}

 The velocities of the components, derived from multiple Gaussian fits
to the profiles, are listed in Table~\ref{art2tab-velIII}.
The mean heliocentric velocities of the components III.b and III.c are
are about +30 and +70~km~s$^{-1}$ respectively.
After correction for the system velocity of +50~km~s$^{-1}$, the
velocities in the frame of the system are -20 to +20~km~s$^{-1}$.
Since the velocities of the emission peak and of the deepest
point of the absorption of a P-Cygni profile are not correlated
(see e.g. Castor and Lamers \cite{art2caslam}) the agreement between these
two values can {\it not} be taken as an indication of the terminal
velocity of the wind. The terminal velocity of the wind can be considerably
higher than the velocities measured from the peak of the emission
and the deepest point of the absorption.
The terminal velocity of the wind can be derived from the extent of the
blue absorption edge. This is about -50~km~s$^{-1}$ heliocentric velocity,
which indicates a terminal velocity of the wind of about 100~km~s$^{-1}$, i.e.
about the same as measured from the Balmer profiles.

The profiles show the presence of an extra absorption component at
-53~km~s$^{-1}$ in the frame of the system (component III.a in
Fig.~\ref{art2fig-lines1}).
This component is probably formed in the wind as the P-Cygni profiles
of the Balmer lines also showed discrete absorption components.

 For those lines of type III that do not show emission, the depth of the
 III.b absorption component is larger than the depth of the III.a absorption
 component. However
 when there is emission in the feature, the III.a component is stronger than
 the III.b component. This suggest that the emission fills in the III.b
 component and reverses the ratio of the strength of the two
 absorption components. In section~\ref{art2sec-IV} we will see that
the UV lines show the same characteristics.

 We conclude that the optical lines of neutral and singly ionized metals
 (type III) show profiles which consist of a P-Cygni profile,
 formed in a wind with a terminal velocity of about 100~km~s$^{-1}$.

 \subsection{Type IV: Ultraviolet absorption lines from
                low-excitation levels
                of neutral and singly ionized metals}
 \label{art2sec-IV}

 In an extensive study of the high-resolution
 IUE spectrum of HD~101584 between 2500 and 3000~\AA,
 229 absorption features have been identified whereas
 41 features remained unidentified (Bakker \cite{art2bakkerart1}).
 The main conclusions are: \newline
 (i) the spectrum of HD~101584
 has the same absorption features as the F0I star
 $\alpha$~Lep, but the lines are intrinsically much broader, \newline
 (ii) the absorption lines are not symmetric, but have
 an excess of {\it red} absorption (this in contradiction
 to lines which are formed in a wind which have a profiles with
 additional blue absorption) and \newline  (iii)  the spread of
 15~km~s$^{-1}$ of the Doppler velocities of the cores of the absorption
 profiles is a factor five larger than measured for the comparison star
 $\alpha$~Lep (Table~\ref{art2tab-velIV}).

 \begin{table}
 \caption{Radial velocities of type IV lines}
 \label{art2tab-velIV}
 \centerline{\begin{tabular}{llll}
 \hline
 Object       & Date      & $v_{\odot}$  &$N$ \\
 \hline
              &           &              &    \\
 HD~101584    & Feb. 1989 & 17\mpm15     & 12 \\
 HD~101584    & Feb. 1990 & 21\mpm8      & 19 \\
              &           &              &    \\
              & Average   & 20\mpm11     & 31 \\
              &           &              &    \\
 $\alpha$ Lep & Dec. 1978 & 22\mpm3      & 19 \\
              &           &              &    \\
 \hline
 \end{tabular}}
 \end{table}

 From the list of UV identifications by Bakker (\cite{art2bakkerart1})
 a selection has been made of the  FeII lines
 which are not significantly
 blended by other lines, and have a secure identification
 (quality factor $Q=3$).
 To prevent the confusion with interstellar
 absorption lines all lines from the ground level,
 $\chi=0.0$~eV, were excluded for this study.
 The 19 selected FeII line are listed in Table~\ref{art2tab-fe}.

 In this paper we show that the
 radial velocity of an absorption line of each ion is correlated with
 the line strength parameter $X$ defined as:

 \begin{equation}
 \label{art2eq-logx}
  \log X=\log gf - 5040 \frac{\chi}{T_{\rm wind}}
 \end{equation}

This parameter is proportional to the expected column density of the ions
(in LTE). Here we will use it as
a measure of the optical depth ($\tau$) of a line.
{}From a comparison between the HIRES spectrum of HD~101584
and $\alpha$~Lep no differences in the occurrence of spectral
features was found and we adopt the photospheric
temperature of a F0Ib star ($\alpha$~Lep)
and for the temperature of the expanding photospheric material,
$T_{\rm wind}=7700$~K.

\begin{figure*}
\centerline{\hbox{\psfig{figure=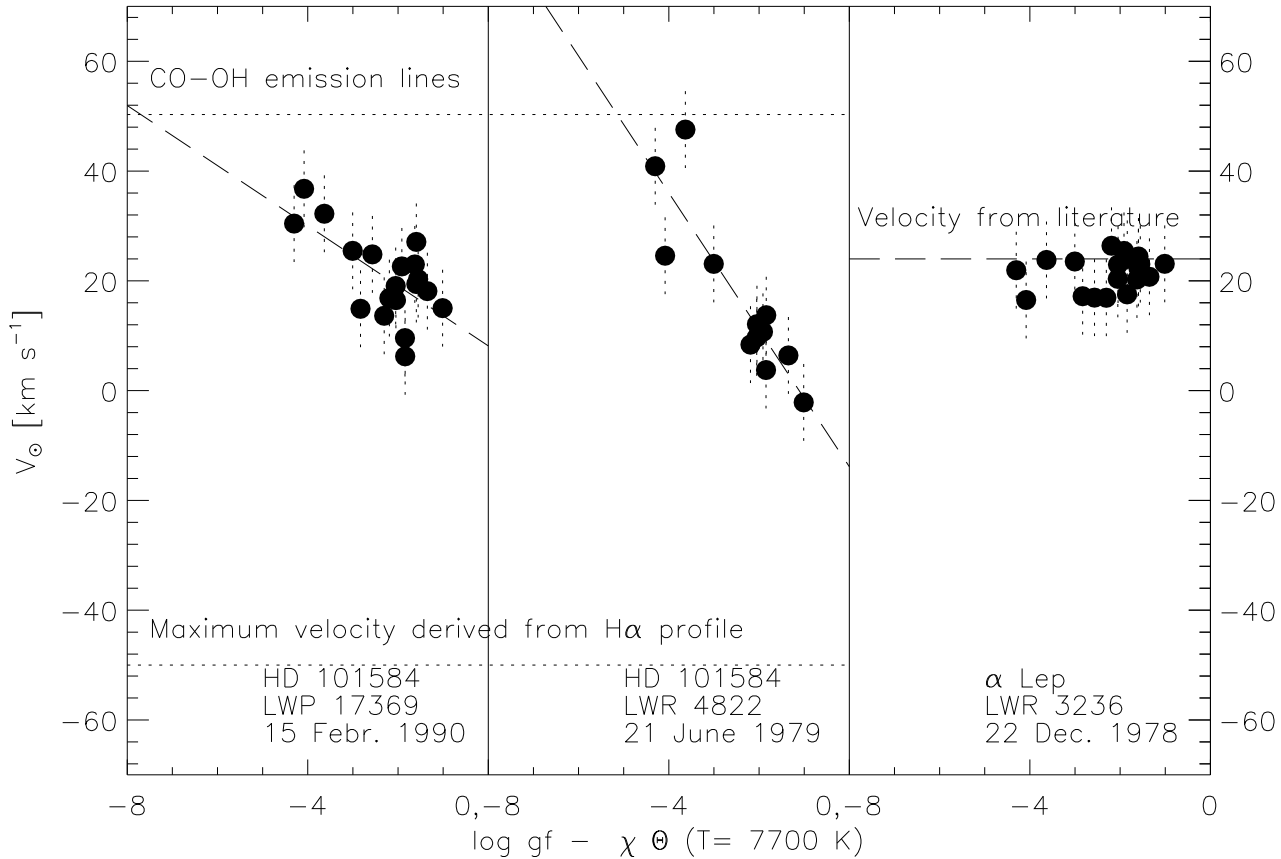,width=\textwidth}}}
\caption{The correlation between $\log X$ and radial velocity $v_{\odot}$
         of UV absorption lines in the spectrum of
         HD~101584 for two different observation dates.
         The slope and offset from stellar velocity
         change in time. The same
         graph has been made for the reference star $\alpha$~Lep.
         There is no gradient found in this star}
\label{art2fig-timegrad}
\end{figure*}

Fig.~\ref{art2fig-timegrad} shows the correlation
between the $\log X$ and the radial velocity for the FeII lines
at two different epochs.
The mean linear relations between the radial velocity and $X$ are

\begin{equation}
\label{art2eq-rel1}
v_{\odot}({\rm FeII}) =  -5.48 \times \log X(T_{\rm wind}=7700 {\rm ~K})
+  8.15
\end{equation}

\noindent
on February 15 1990, and

\begin{equation}
\label{art2eq-rel2}
v_{\odot}({\rm FeII}) = -12.45 \times \log X(T_{\rm wind}=7700 {\rm ~K})
- 13.47
\end{equation}

\noindent
on June 21 1989.

The radial velocity found for $\alpha$~Lep of 22\mpm3~km~s$^{-1}$
is in agreement with the radial velocity given
by Anderson~{\sl et al.} (\cite{art2anderson}) of 24.3~km~s$^{-1}$.

There is clearly a correlation between the $\log X$ and the radial
velocity.  Other ionization degrees and other species (CrII, MnI etc.)
show the same correlation but with slightly different slopes and offsets.
We have made the same test on type III lines, i.e. the low-excitation
metal lines in the optical,
and also found a correlation between $\log X$ and the radial velocity.

 We can explain the correlation with a simple model in which
 the low-excitation UV and optical lines are formed in the wind
 (Bakker \cite{art2bakkerlhl}).
 The stronger the line strength parameter $\log X$
 the further out in the wind the line is formed.
 The wind accelerates outwards and  we expect to see all wind lines to
 be blue-shifted with respect to the stellar velocity, but
 the stronger the line, the larger the blue-shift will be.

 In this paper we propose a different explanation for the correlation
 between the line strength and the radial velocity using
 the results obtained from the optical P-Cygni lines
 (see Fig.~\ref{art2fig-lines1} type III), but with the same
conclusion that the UV lines are formed in the wind.
 These optical lines
 have the same excitation conditions and are from the same species as
 the UV absorption lines.
The optical P-Cygni lines have multiple components.
We suggest that the P-Cygni lines in the UV also have multiple
components, but that these are not resolved
due to a lower resolution of the IUE observations.
We only see
 one broad absorption with the emission component being lost in the continuum.
The separation between the two absorption components in the optical lines is
33~km~s$^{-1}$
and the resolution of the IUE spectra is 30~km~s$^{-1}$. So the separation
is expected to be barely detectable in the UV lines.
A careful examination of the HIRES UV spectra
 shows that most of the absorption lines indeed show a hint of a second
 component.
 This explains the anomalies found in the UV spectrum: \newline
 \indent
 {\sl (i) Large spread in observed radial velocities.}
 The correlation between $\log X$ and radial velocities introduces
 a large spread
 in the average velocity of the UV lines. The average radial velocity
 of $20\pm11$~km~s$^{-1}$ is in good agreement with the velocity predicted
 for a blend between III.a and III.b lines. \newline
 \indent
 {\sl (ii) Large equivalent width.}
 As the absorption lines are the result of two separate absorption components
 we observe a large equivalent width. Also because of the velocity
 stratification the absorption is more efficient and optical thickness
 is reached at a higher column density. \newline
 \indent
 {\sl (iii) Line asymmetry: too much red absorption.}
  We have seen in the optical spectrum that for the weaker lines the
 III.b absorption
 component is stronger than the III.a component.
 When the emission is important it
 fills in the III.b component and the III.a component dominates the feature.
 In fitting a Gaussian to the feature we observe
 the red component as additional  absorption.

 From Fig.~\ref{art2fig-timegrad} we learn that
 the slope and offset of the fit for HD~101584
 changes in time (Eq.~\ref{art2eq-rel1} and \ref{art2eq-rel2}).
This means that the two absorption components
 move relative to each other or that the emission varies in strength.

The absorption components of the UV P-Cygni profiles extent to a
heliocentric velocity of about -50~km~s$^{-1}$ or -100~km~s$^{-1}$
in the velocity frame of the star (Fig.~\ref{art2fig-lines1}).
This indicates a terminal
velocity of the wind of the star of about 100~km~s$^{-1}$,
i.e. the same value
as derived from the P-Cygni profiles in the optical.

 \subsection{Type V: Optical pure absorption lines from
                         low-excitation levels of neutral metals}
 \label{art2sec-V}

The optical spectra show pure absorption lines from
FeI which deviate in shape from all the other absorption features
(Fig.~\ref{art2fig-lines1} type V). The
features are much narrower in shape and do not show emission components.
The small width of the features and their low-excitation
suggest that these lines are formed
in a colder region than the other absorption lines.

 \begin{table}
 \caption{Radial velocities of type V lines}
 \label{art2tab-velV}
 \centerline{\begin{tabular}{lll}
 \hline
 Date      & $v_{\odot}$  &$N$ \\
 \hline
           &              &    \\
 Feb. 1989 & 50.5\mpm2.1  & 02 \\
 Jan. 1990 & 50.3\mpm1.7  & 03 \\
 Apr. 1992 &              & 00 \\
 Feb. 1993 &              & 00 \\
           &              &    \\
 Average   & 50.4\mpm1.3  & 05 \\
           &              &    \\
 \hline
 \end{tabular}}
 \end{table}

The radial velocities of these lines are listed in Table~\ref{art2tab-velV}.
The mean radial velocity is 50.4\mpm1.3~km~s$^{-1}$ (heliocentric) which is
exactly equal to the system velocity derived from the CO and OH lines.
The radial velocity of the lines does not vary in time.
 On the basis of their radial velocity, their narrow width and
 the range of excitation
 levels ($0.0\leq \chi \leq 0.22$~eV) we argue that these lines are of
 circumsystem origin and that the gas temperature is lower than the
 wind and photospheric temperatures, maybe as low as
 $T_{\rm gas} \approx 10^{3}$~K.

 The only stable geometry which can sustain this material
 is a (Keplerian) disk with an inclination angle small enough for the
 line-of-sight to the star to pass through disk material.
 With a temperature of about $T_{\rm gas} \approx 10^{3}$~K, the
line forming region coincides with the inner region of the dust,
i.e. at about $20 R_{\ast}$ where $T_{\rm dust}=1240$~K.

 The NaI~D1 and D2 profiles show a narrow absorption
 component at a heliocentric velocity of -32 + 2.1~km~s$^{-1}$,
and a velocity in the frame of the system of -82~km~s$^{-1}$.
This velocity is rather high for interstellar lines
and the same as the edge velocity of the \ha line.
Therefore the NaID lines are most likely
formed far out in the stellar wind of HD~101584.
This implies that
we have not detected a single interstellar line in the optical
spectrum of  HD~101584 !

\subsection{Type VI: Optical pure emission lines from
                     low-excitation levels of neutral and singly
                     ionized metals}

\label{art2sec-VI}

 The pure emission lines in the optical spectrum are all from neutral and
 singly ionized metals like e.g. FeI, TiI, MnI, NiI and TiII, and are hardly
 resolved in the CAT/CES spectra.
(Table~\ref{art2tab-optical}  shows that the
emission lines at 4334~\AA~ and 7451~\AA~ have tentatively been identified
with LaII and YII respectively. However this identification is uncertain
and the lines may be due to unknown metal lines. If this
identification is confirmed the presence of these s-elements would pose
strict constraints on the evolutionary status of HD~101584).

\begin{table}
\caption{Radial velocities of type VI lines}
\label{art2tab-velVI}
\centerline{\begin{tabular}{lll}
\hline
Date      &$v_{\odot}$   &$N$\\
\hline
          &              &   \\
Feb. 1989 & 48.1\mpm0.4  &16 \\
Jan. 1990 & 51.0\mpm3.0  &01 \\
Apr. 1992 & 51.1\mpm0.4  &14 \\
Feb. 1993 & 50.5\mpm0.6  &12 \\
          &              &   \\
Average   & 49.8\mpm0.6  &43 \\
          &              &   \\
\hline
\end{tabular}}
\end{table}

The $FWHM$ velocity of the emission lines is 25~km~s$^{-1}$.
 The average heliocentric velocity of the pure
 emission lines on the four epochs is 49.8\mpm0.6~km~s$^{-1}$
 (Table~\ref{art2tab-velVI}). Within the
 error  this is equal to the system velocity of 50.3\mpm2.0~km~s$^{-1}$
derived from the lines of CO and OH.
 It was already noted by  Trams~{\sl et al.} (\cite{art2tramsetal1})
 that the  radial velocity
 of the pure emission lines does not vary in time. They
 argue that these lines are formed  either in blobs which were ejected
 almost perpendicular to the line of sight, or more likely in a
 distant ring  or flattened shell which is tilted towards the observer.
 We can now exclude the blob interpretation, because we have shown
that the same velocity is seen in optical {\it absorption} lines.
We conclude that the narrow optical emission lines are formed in the same
circumsystem disk as  the optical absorption lines
of low-excitation metals. This disk must be seen almost edge-on to account
for the low-excitation absorption lines.

The $FWHM$ of 25~km~s$^{-1}$ of the unresolved emission lines
gives a lower limit on the
distance between the disk and the star from its Keplerian velocity.
If we assume that the mass of the star or of the binary system
is of the order of 1 or 2~\msol, we find that the emission lines are
formed at a distance of at least 3~AU. Assuming a stellar radius
of $R_{\ast} = 38$~\rsol for a 0.54~\msol post-AGB stars
(Sect.~\ref{art2sec-nature}) we find that the distance
$r_{\rm disk} \simeq 20 R_{\ast}$.
This distance is the same as the distance found for the inner dust
radius derived from the infrared excess (Sect.~\ref{art2sec-energy}).

 \subsection{Type VII: Infrared emission feature}
 \label{art2sec-VII}

\begin{figure}
\centerline{\hbox{\psfig{figure=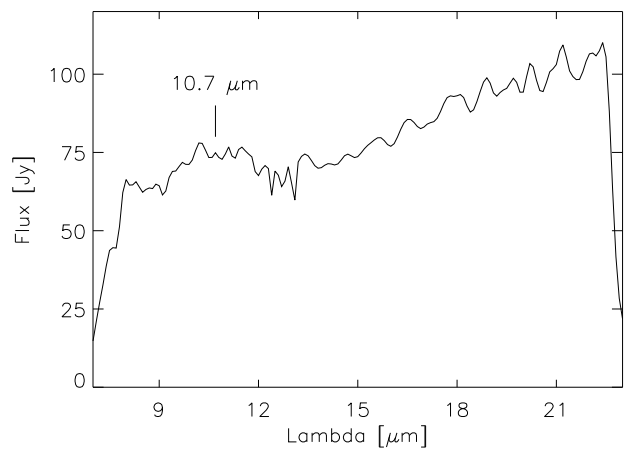,width=\columnwidth}}}
\caption{Low-resolution IRAS spectrum of HD~101584. Notice the
         detection of a weak
         10.7~$\mu$m infrared feature in emission}
\label{art2fig-lrs}
\end{figure}

 The IRAS low-resolution spectrum (LRS, Fig.~\ref{art2fig-lrs}) shows
 an  infrared emission feature. We have fitted this feature with
 a Gaussian and found a central wavelength of 10.7~$\mu$m.
 The 10~\mum feature is attributed to oxygen rich material.
 This, together with
 the non-detection of a 11.3~$\mu$m feature and the detection of
 OH maser emission indicates that the circumstellar environment is oxygen
 rich.

 \subsection{Type VIII: Molecular line emission}
 \label{art2sec-VIII}

 HD~101584 shows  molecular line  emission of $^{12}$CO($J=1-0$)
 (2.6~mm) and  $^{12}$CO($J=2-1$)  (1.3~mm)
 (Trams~{\sl et al.} \cite{art2tramsetal1}; Loup~{\sl et al.}
\cite{art2loupetal}),
 vibrational-rotational bands heads of
 $^{12}$CO($v''=2-0$) (2.29~$\mu$m), $^{12}$CO($v''=3-1$)
 (2.32~$\mu$m),
 $^{12}$CO($v''=4-2$) (2.35~$\mu$m) (Oudmaijer~{\sl et al.}
 \cite{art2oudmaijernir}) and OH maser emission at 1667~MHz
 (Te Lintel Hekkert~{\sl et al.} \cite{art2lintelhekkertetal}).
 The CO millimeter emission originates from
 material which is relatively far away
 from the star ($ r \geq 5\times 10^{4}~R_{\ast}$)
 with temperatures of several tens of K.
 The OH 1667~MHz main line is formed closer to the star with temperatures
 on the order of $T_{\rm dust} \approx 150-280$~K
 (page 247 in Elitzur \cite{art2elitzur}) at
 a distance of the order of $r_{\rm OH} \approx 3 \times10^{3}~R_{\ast}$.

 Both CO and OH are integrated over a large volume
 of emitting material and any asymmetry in the shape or velocity
 of the emitting volume of material is smeared out in
 the profile. For this reason we assume that the symmetry axis
 of the CO and OH line profiles is the radial velocity of the system.
 Both CO and OH yield the same system velocity with an average value of
 50.3\mpm2.0~km~s$^{-1}$.

CO  shows several emission peaks
 symmetrically around this velocity at a relative velocity to
the system of 0, 43 and 130~km~s$^{-1}$. The contribution from the
circumsystem disk should be at the system velocity with a
width of $\approx 25$~km~s$^{-1}$. We identify the central
component of the CO emission lines profile as due to the
circumsystem disk. The features at 130~km~s$^{-1}$ is close to
the terminal velocity derived from \ha and we identify this
component as due to recent mass-loss. The component at 43~km~s$^{-1}$
is due to an earlier mass-loss episode of the star
(AGB) when the star had a larger radius, a lower escape
velocity and thus a lower terminal velocity of the wind.

The CO emission in the near-infrared originates from hot CO gas
($T_{\rm gas} \approx 2\times10^3$~K).
This temperature is of the same order as
the dust temperature in the circumsystem disk which suggests
that the near-infrared CO emission originates
from the circumsystem disk.

The OH maser 1667~MHz shows a double  peaked emission profile with
peaks at relative velocity to the system of $\approx$ 30~km~s$^{-1}$. As this
velocity does not corresponds with a CO peak it is unclear where
this OH maser originates.

 \subsection{Summary}

 The wealth on information on the radial velocities and shapes
 of spectral
 lines of HD~101584 gives clues about the  nature and
 geometry of the system.
 Table~\ref{art2tab-vel} gives a summary of the
 average heliocentric velocities and the velocities in the frame of the system,
  i.e. relative to the velocity of the CO and OH lines
 for each category of spectral lines.

 \begin{table*}
 \caption{Heliocentric radial velocities of different categories
 of spectral lines in the UV and optical spectrum of HD~101584}
 \label{art2tab-vel}
 \centerline{\begin{tabular}{llrrl}
 \hline
 \multicolumn{2}{c}{Category} &$v_{\odot}$&$\delta v$& Remark            \\
      &                       &\multicolumn{2}{c}{[km~s$^{-1}$]}&        \\
 \hline
      &                           &            &          &
     \\
 I.a  &high $\chi$, optical       & 24.7\mpm1.7& -25\mpm3 &abs., only NI(3) at
Apr. '92 \\
 I.b  &                           & 56.9\mpm5.9&   7\mpm7 &abs., B8-9I-II star
     \\
 II.a &Balmer  lines              &-58\mpm2    &-108\mpm3 &abs., shell
component     \\
 II.b &                           &-10\mpm4.9  & -60\mpm5 &abs., shell
component     \\
 II.c &                           & 47\mpm4.9  &  -3\mpm3 &abs., wind and
photosphere  \\
 II.d &                           & 81\mpm1.5  & +33\mpm3 &em.,  wind in
emission    \\
 III.a&low  $\chi$, optical       &-3.1\mpm1.8 & -53\mpm3 &abs., shell
component     \\
 III.b&                           & 30.2\mpm4.8& -20\mpm5 &abs., wind component
     \\
 III.c&                           & 69.6\mpm4.2&  20\mpm5 &em.,  wind component
     \\
 IV   &low  $\chi$, UV            & 20\mpm11   & -30\mpm11&abs., wind component
     \\
  V   &narrow, low $\chi$, optical& 50.4\mpm1.3&   0\mpm2 &abs., at system
velocity  \\
 VI   &narrow, low $\chi$, optical& 49.8\mpm0.6&   0\mpm3 &em.,  at system
velocity  \\
 VII  &10~$\mu$m AGB remnant      &            &          &em.,  in dusty
regions    \\
 VIII &CO and OH AGB remnant      & 50.3\mpm2.0&   0\mpm2 &em.,  at system
velocity  \\
      &                           &            &          &
     \\
 \hline
 \end{tabular}}
\centerline{\( \delta v = v_{\odot} - v_{\odot {\rm VIII}} \)}
 \end{table*}

 \section{The nature of HD~101584}
 \label{art2sec-nature}

 In this section  we will make a synthesis of the results from
 the study on the energy distribution and the study on the spectral lines.

\subsection{Summary of the observational characteristics}

\begin{figure*}
\vspace*{-3cm}
\centerline{\hbox{\psfig{figure=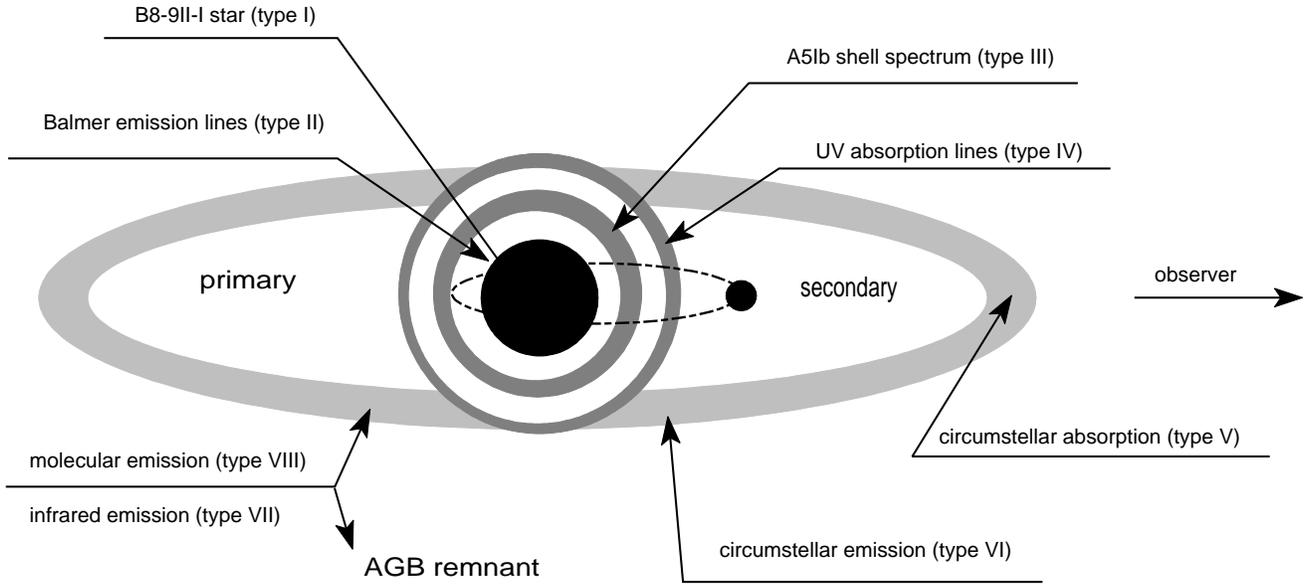,width=\textwidth}}}
\caption{A model for HD~101584}
\label{art2fig-mode}
\end{figure*}

 The observations can be summarized as follows and can be visualized
 with the help of Fig.~\ref{art2fig-mode}:

 1. The CO and OH emission lines show that the system has a heliocentric
velocity of + 50\mpm2~km~s$^{-1}$.

\begin{figure*}
\centerline{\hbox{\psfig{figure=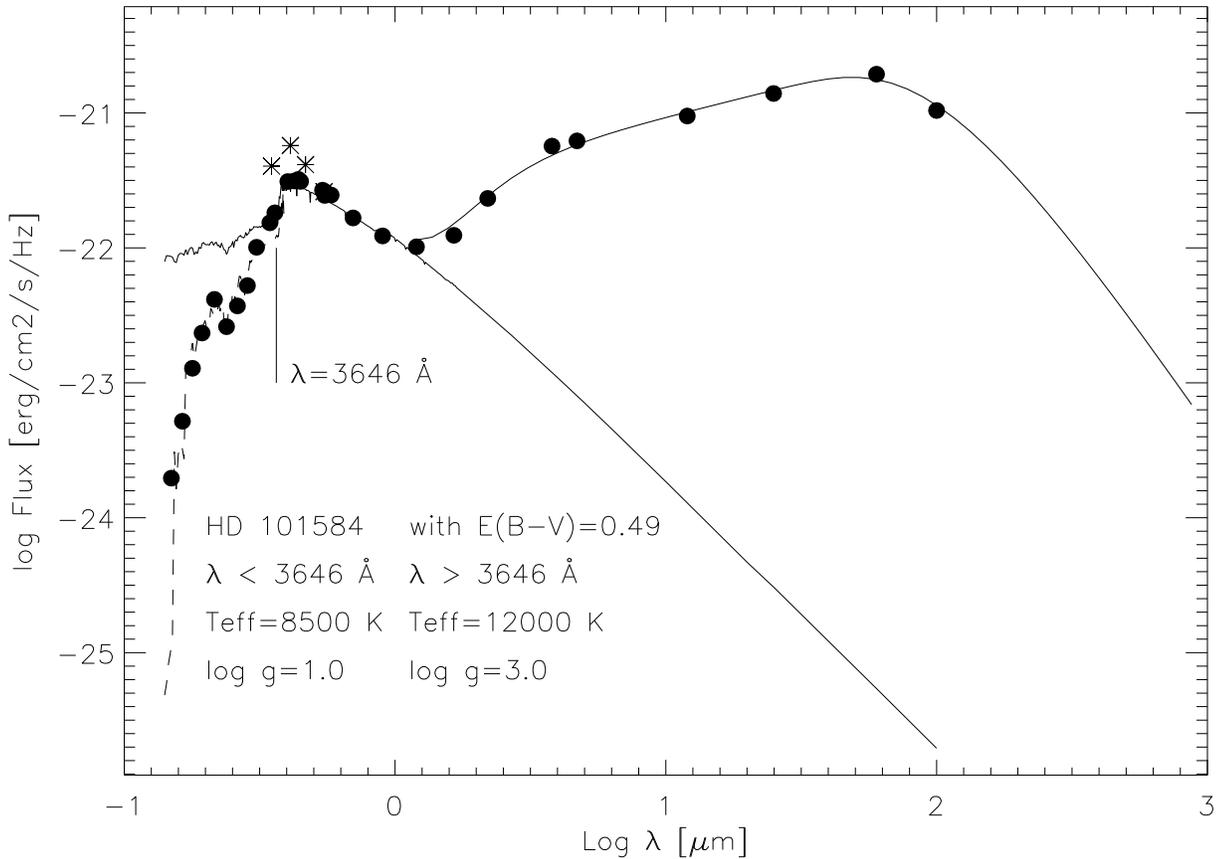,width=\textwidth}}}
\caption{A B9II (solid line) and a A5I (dashed line, only for
$\lambda \leq 3646$~\AA~)  Kurucz models are fitted to the dereddened
energy distribution ($E(B-V)=0.49$) of HD~101584.
Due to the many wind lines in the
blue and UV part of the spectrum, we observe an
iron curtain in front of the B-star and the UV SED mimics an A to F-type
star blue-wards of the Balmer jump ($\lambda = 3646$~\AA)}
\label{art2fig-sedab}
\end{figure*}

2. The system contains a late type B-star which produces the high-excitation
absorption lines in the optical spectrum.
The Geneva photometry shows that the star has $T_{eff}=12000\pm1000$~K
and $\log g =3.0\pm1.0$, so it is a late-B star of type about
B9. With this gravity the star would be of luminosity class II
(see also Fig.~\ref{art2fig-sedab}).

3. This indicates an extinction of $E(B-V)=0.49\pm0.05$.

4. The high-excitation photospheric absorption lines show radial velocity
variations with an amplitude of at least 10~km~s$^{-1}$, which suggests
that the star has a companion. We will refer to the late-B star as
``the primary''. The present data are too scarce to determine a period.
However Bakker {\sl et al.} (\cite{art2bakkerart5}) claim  the presence
of a 218~day photometric period, and argue that this period is
likely present in the Doppler velocities of the high-excitation absorption
lines. This strongly support the binary scenario.
We did not find any spectroscopic signature of the secondary. This
suggests that its visual luminosity would be considerably
fainter than that of the primary by at least about a factor 50,
which is not exceptional for the companion of a supergiant.

5. The Balmer lines and  the low-excitation metal lines in the optical
and in the UV show P-Cygni profile. This indicates the presence
of a strong stellar wind. This wind is most probably due to the primary
because we do not see any spectroscopic evidence for the secondary.
The extent of the absorption wings
of the P-Cygni profiles of the Balmer lines and the metal lines
all indicate a terminal wind velocity, \vinf, of about 100\mpm30~km~s$^{-1}$.

6. The UV spectrum is dominated by the enormous number of
metal lines, most of which show P-Cygni profiles. This makes
the UV spectrum of HD~101584 to resemble the spectrum of a F-type supergiant
although the star has a late-B spectral type (we see an iron
curtain in front of the B-star).
In fact the UV absorption lines are the same as those observed in
the F0Ib supergiant $\alpha$~Lep, but the lines are
stronger in the spectrum of HD~101584 (Bakker \cite{art2bakkerart1}).
The presence of many P-Cygni profiles and blue-shifted
absorption profiles from low-excitation metal lines in the {\it optical}
spectrum explains the original classification of the star as
F0Iape. Fig.~\ref{art2fig-sedab} displays a 12000~K (B9II) and a 8500~K
(A5I) Kurucz model fitted to the dereddened ($E(B-V)=0.49$)
energy distribution of HD~101584. For wavelengths smaller than
the Balmer jump the
SED resembles a cool A star, while for wavelengths larger than the
Balmer jump the SED resembles a hot B star.

7. The Balmer lines and the wind lines in the optical and in the UV
show additional absorption components superimposed on the underlying
P-Cygni profiles. These are qualitatively similar to the discrete
absorption components observed in the P-Cygni profiles of almost all
early type stars. The absorption components in
the P-Cygni profiles of HD~101584 are at velocities relative to the star
of -20 and -53~km~s$^{-1}$ (optical and UV P-Cygni lines) and -108
and -60~km~s$^{-1}$ (Balmer lines).

8. The terminal velocity of the wind can be used to derive
information about the gravity of the star if we neglect the presence of a
companion.
The observations of stellar winds of early type stars and the
theory of radiation driven winds indicate that \vinf is proportional
to the photospheric escape velocity, $v_{esc}$.
The ratio $v_{\infty}/v_{esc}$
is 1.2\mpm0.3 for late-B supergiants (Lamers~{\sl et al.}
\cite{art2lamersetal}). This
implies an escape velocity of 90\mpm30~km~s$^{-1}$.

9. The optical spectrum shows narrow low-excitation {\it absorption}
lines of neutral Fe at the system velocity.
Since these lines are much narrower
than the photospheric high-excitation absorption lines and they do not
show the radial velocity variations, they must be formed in a circumsystem
disk that is in front of the star.

10. The optical spectrum also shows narrow {\it emission} lines from
neutral and
singly ionized gas at the system velocity. These lines are formed in the
circumsystem gas that also
formed the absorptions of the low-excitation FeI lines.

11. The molecular emission of CO in the infrared (vibrational transitions)
shows the presence of circumsystem material with a temperature on
the order of $T_{\rm gas} \approx 2\times10^{3}$~K around the system.
The emission lines from CO at millimeter wavelengths (rotational transitions)
and the OH maser line at 1667~MHz
at the system velocity show that the circumsystem gas extends to large
distances of about $10^{3}$ to $10^{5} R_{\ast}$ at temperatures of
$10^{1}$ to $10^{2}$~K.

12. The huge infrared excess can be energetically supported by
the star if we assume that it has a
reddening of $E(B-V)=0.38\pm0.2$ due to  a spherically symmetric dust shell.
{}From the IR excess we find that the dust is typically at a distance
of 20 to $10^{5} R_{\ast}$. As the system is seen edge-on the
observed reddening will be larger than $E(B-V)=0.38$. This extinction implies
that the star has a spectral type earlier than about A2.

13. The fact
that we see the low-excitation FeI lines formed in the circumsystem disk,
in {\it absorption}, means that the line-of-sight to the primary passes
through the circumsystem disk. So the disk must be observed almost edge-on.
We can then use the width of the
circumsystem emission lines of low-excitation metals to estimate the
distance from the star. Assuming that the binary system has a total mass
of about 1 or 2~\msol (see below), the $FWHM$
of 25~km~s$^{-1}$ of the emission
indicates a lower limit on the distance of $20 R_{\ast}$.
This is of the same order as the dust inner radius.

Fig.~\ref{art2fig-mode} shows
a simple cartoon of the model of geometry of HD~101584 proposed. This model
explains all different categories of spectral lines in a consistent way.
We ordered the spectral features in decreasing excitation conditions.
This is reflected in the model. The higher the excitation level
of the transition, the closer it originates to the central B-star.
However this model does not say anything about the evolutionary
status (nature) of HD~101584. In the next sections we will
use the observations and the model to derive the
mass and the nature of HD~101584.

\subsection{The mass, luminosity and distance}

The stellar wind from HD~101584 has a low terminal velocity of only
about 100\mpm30~km~s$^{-1}$.
This suggests an escape velocity of 90\mpm30~km~s$^{-1}$.
In Table~\ref{art2tab-popi} we give expected parameters for Pop I
late-B supergiants and compare these with the data of HD~101584.

The relation between mass and luminosity at $T_{eff}=12000$~K was derived
from the evolutionary tracks of
Schaller~{\sl et al.} (\cite{art2schaller}) under the assumption that
the star is moving to the right in the Hertzprung Russell diagram after
it left the main sequence.
The radius was derived from $L$ and $T_{eff}$.
The correction factor for the effective gravity due to radiation
pressure by electron scattering is $\Gamma = 2.66\times10^{-5} L/M$
when $L$ and $M$ are in solar units and hydrogen is supposed to be
almost fully ionized in the atmosphere. This is a reasonable
approximation for late-B supergiants (Lamers~{\sl et al.}
\cite{art2lamersetal}).

The escape velocities derived in this way are considerably
larger than the observed value of
about 90~km~s$^{-1}$. This suggests that either the star is even more
luminous or that the star is in a later evolutionary phase, and it is
evolving to the blue in the HRD after having been a red supergiant.

The distance of the star, derived from $L$, from the bolometric correction of
$BC\simeq  -0.70 \pm 0.05$, from $E(B-V)=0.49\pm0.05$ and from
$V= 6.95$ is also listed.
The distance is between 4 and 11~kpc, which implies a height
above the galactic plane between  0.4 and 1.1~kpc.

\begin{table*}
\caption{Stellar and wind parameters for population I stars}
\label{art2tab-popi}
\centerline{\begin{tabular}{lllllllll}
\hline
Pop&$M_{i}$&$\log L$&$M$    &$R$    &$\Gamma$&$v_{esc}$&$D$&$v_{\infty}$\\
   &[\msol]&[\lsol] &[\msol]&[\rsol]&        &[km~s$^{-1}$]&[kpc]&[km~s$^{-1}$]
    \\
\hline
     &       &        &       &       &        &         &     &    \\
I    &60     &6.00    &49.2   &230    &0.47    &210      &10.3 &227 \\
I    &50     &5.88    &43.3   &200    &0.40    &220      &9.0  &238 \\
I    &40     &5.69    &36.0   &160    &0.31    &240      &7.1  &259 \\
I    &30     &5.47    &28.0   &130    &0.24    &250      &5.4  &270 \\
I    &20     &5.10    &19.0   &80     &0.15    &280      &4.1  &302 \\
     &       &        &       &       &        &         &     &    \\
\hline
\end{tabular}}
\end{table*}

Table~\ref{art2tab-pagb} gives a similar table in the case
that HD 101584 is a post-AGB star. In this case we derive the luminosity
from the mass-luminosity relation of Boothroyd and Sackmann
(\cite{art2bootsack}).
The rest of the data are derived in the same way as for Pop I stars.

\begin{table*}
\caption{Stellar and wind parameters for post-AGB stars}
\label{art2tab-pagb}
\centerline{\begin{tabular}{lllllll}
\hline
$\log L$&$M$    &$R$    &$\Gamma$&$v_{esc}$ &$D$  &$v_{\infty}$\\
{}~[\lsol]&[\msol]&[\rsol]&        &[km~s$^{-1}$]    &[kpc]&[km~s$^{-1}$]  \\
\hline
        &       &       &        &          &     &        \\
4.10    &0.70   &26     &0.43    &77        &1.1  &83      \\
4.00    &0.65   &23     &0.37    &83        &1.0  &90      \\
3.87    &0.60   &20     &0.29    &90        &0.86 &97      \\
3.70    &0.55   &16     &0.22    &100       &0.71 &108     \\
3.52    &0.52   &13     &0.15    &114       &0.59 &123     \\
        &       &       &        &          &     &        \\
\hline
\end{tabular}}
\end{table*}

Tables~\ref{art2tab-popi} and \ref{art2tab-pagb}
show that an escape velocity of $v_{esc}=90\pm30$~km~s$^{-1}$
is only consistent with Post-AGB stars. The predicted values of
$v_{esc}$ of massive stars are about a factor two to three too high.
This suggests that HD~101584 is a post-AGB star
at a distance between 0.6 and 1.1~kpc. At that distance the star
is between 60 and 100~pc above the Galactic plane.
The low-mass nature of the progenitor of HD~101584 is in agreement with the
galactic latitude and with the high space velocity.

In this discussion we have neglected the role of the companion
in the determination of the mass-loss rate and the terminal velocity
of the wind. The presence of a circumsystem disk shows that the
companion plays or has played a role in the mass-loss history
of the star.  So the companion may influence the stellar wind,
and reduce the effective gravity of the primary and therefore decrease
the terminal velocity and increase the mass-loss rate. If this
is the case, the low terminal velocity cannot be used as an argument against
a massive primary.

\subsection{Is HD~101584 a post-AGB binary?}

{}From studies of bright post-AGB stars Waters~{\sl et al.}
(\cite{art2watersserena})
found that the occurrence of a near-infrared excess is
strongly correlated with binarity.
 The infrared-excess of HD~101584  has a local minimum at 1.25~$\mu$m,
which indicates that the circumsystem dust has two components of
different temperatures.
 The total infrared energy distribution can be interpreted  as
 the combination of a near-infrared excess
 with a dust temperature of $T_{\rm dust} \approx 1240$~K and a far-infrared
 excess due to an expanding dust remnant.
Dust around a central star of about 12000~K will reach a
temperature of 1240~K at an inner dust radius of about
$10^{2} R_{\ast}$.

If HD~101584 is indeed a low-mass post-AGB star it differs from the
other post-AGB binaries in that its dust is O-rich. This can be inferred
from the fact that the star shows OH maser lines and the 10~$\mu$m
infrared emission feature. Other
post-AGB binaries with a dusty disk, such as  HR~4049, HD~213985 and the
Red Rectangle (HD~44179), are all surrounded by C-rich dust.

The mass-loss rate of HD~101584 can be estimated by
comparing its stellar wind with that of the B8Ia star
$\beta$~Ori which has a mass-loss rate of \mdot$\simeq 10^{-7}$~\msol~\pyr~
and $v_{\infty} \simeq 350$~km~s$^{-1}$. The UV spectrum of HD~101584
is similar to that of $\beta$~Ori, but it shows more blue-shifted
UV absorption lines of singly ionized metals
(Lamers~{\sl et al.} \cite{art2lamstal}, Lamers~{\sl et al.}
\cite{art2lamersetal}).
This means that the column density of the stellar wind of HD~101584 is
higher or of the same order of magnitude as that of $\beta$~Ori.
The column density of a stellar wind scales as
$N \sim$\mdot$/\left( v_{\infty} R_{\ast} \right)$,
so \mdot$\sim N v_{\infty} R_{\ast}$.
Assuming R$_{\ast}=150$~\rsol for $\beta$~Ori and 20~\rsol for
HD~101584, and $v_{\infty}=350$~km~s$^{-1}$ and 100~km~s$^{-1}$
for the two stars
respectively we find a rough estimate of
\mdot(HD~101584)$\geq 4 \times 10^{-9}$~\msol~\pyr.

The large mass-loss rate of HD~101584, as inferred from the
many wind lines, is abnormal for post-AGB stars
and suggests a very low effective gravity. This would be consistent
with a massive post-AGB star of about 0.7~\msol
(see Table~\ref{art2tab-pagb}). However the evolutionary transition time
between AGB and Planetary Nebula of a 0.7~\msol post-AGB star
is about $10^{2}$~years (Bl\"{o}cker~\cite{art2blocker}).
This contradicts the fact that the star has not changed its spectral
type and visual magnitude in the last 20 years
systematically (Humpreys and Ney \cite{art2humphreysney}).
Therefore HD~101584 is probably not a massive post-AGB star.

If HD~101584 is a low-mass post-AGB star of
0.5~\msol$\leq M_{\ast} \leq 0.6$~\msol
the higher mass-loss rate and the low effective gravity could be
due to
the presence of a nearby companion. A decrease of the
effective gravity results in an increase of the mass-loss
rate and a decrease of the terminal velocity $v_{\infty}$.

We conclude that HD~101584 is probably a post-AGB star
(based on its high galactic latitude), of intermediate-mass
(based on the fact that it has not shown systematic changes in
$T_{eff}$ and $M_{v}$) with a nearby companion that affects the
stellar wind (based on the high mass-loss rate and low $v_{\infty}$).

\section{Conclusions}
\label{art2sec-conc}

 We have studied the complete energy distribution
 (1190~\AA~ to 100~$\mu$m) and  UV, optical and infrared spectra of
 the enigmatic object HD~101584 and made a consistent
 model of the geometry of HD~101584 (Fig.~\ref{art2fig-mode})
 and the nature of the observed star.
 The system contains a cool B-type post-AGB
 star which probably occurs in a close binary
 system with a white dwarf or a low-mass main sequence star.
 The primary suffers from a high mass-loss rate of the
 order of \mdot$\simeq10^{-8}$~\msol~\pyr.
 The very large number of UV metal lines formed in the wind
 drastically decrease the energy distribution in the UV and
 hides the B-star behind an iron curtain. The UV spectrum mimics
 the photospheric spectrum of a F0I star.
 An upper-limit on the semi-major axis of the system is
 given by the dust inner radius and is of the order of $10^{2} R_{\ast}$.
 The system has a circumsystem disk at a radius of the
 same order as the dust inner radius. This disk is observed by the
 detection of narrow emission and absorption lines.

 \acknowledgements{
 The authors would like to thank Ren\'{e} Oudmaijer and
 Lucky Achmad for reading
 and discussing this paper.
 The author was supported by grant no. 782-371-040 by ASTRON,
 which receives funds from the Netherlands Organization for
 the Advancement of Pure Research (NWO).
LBFMW is supported by the Royal Netherlands Academy of Arts and Sciences.
 This research has made use of the Simbad database, operated at
 CDS, Strasbourg, France, and the IUE database UNSPL at the
 University of Utrecht and VILSPA.  CAT/CES is operated by ESO.
CW thanks the staff of the Geneva Observatory for their kind permission
to use the Swiss Telescope at La Silla.}

\appendix

\section{UV FeII lines}
\label{art2ap-uvFeII}

 The FeII UV lines used in this paper are
 from the UV identification table
 of Bakker (\cite{art2bakkerart1}). All information about the method used
 in determining the  profile parameters can we found in this article.

 \begin{table*}
 \caption{19 selected UV FeII absorption
            lines of HD~101584 and $\alpha$~Lep}
 \label{art2tab-fe}
 \centerline{\begin{tabular}{|lllrrrr|}
 \hline
 $\lambda_{lab.}$ [\AA]   &
 Multiplet  & $\chi $ [eV] &
 $\log gf$& \multicolumn{3}{c|}{$v_{\odot}$ [km~s$^{-1}$] $^{*}$}  \\
 & & & & \multicolumn{2}{c}{HD~101584} & $\alpha$~Lep  \\
 & & & & LWP17369 & LWR4822 &LWR3236 \\
 \hline
          &     &      &       &    &    &    \\
 2509.117 & 242 & 3.23 & -0.89 & 27 & 23 & 24 \\
 2513.372 & 207 & 3.19 & -0.48 & 26 &    & 17 \\
 2533.626 & 159 & 2.65 & +0.14 & 19 &    & 22 \\
 2569.775 & 266 & 3.41 & -1.85 & 37 & 25 & 17 \\
 2619.071 & 171 & 2.79 & -0.48 & 15 &    & 17 \\
 2664.665 & 263 & 3.37 & +0.36 & 10 &  4 & 18 \\
 2714.414 & 063 & 0.98 & -0.37 & 15 & -2 & 23 \\
 2730.735 & 062 & 1.07 & -0.65 & 18 &  6 & 21 \\
 2753.289 & 235 & 3.25 & +0.50 & 24 &    & 20 \\
 2761.813 & 063 & 1.09 & -0.84 & 21 &    & 23 \\
 2767.500 & 235 & 3.23 & +0.52 & 28 &    & 24 \\
 2783.690 & 234 & 3.23 & +0.20 & 24 & 11 & 25 \\
 2880.750 & 061 & 0.98 & -1.20 &  7 & 14 & 24 \\
 2917.465 & 061 & 1.04 & -2.15 & 16 &    & 17 \\
 2926.584 & 060 & 0.98 & -1.41 & 19 & 10 & 20 \\
 2947.658 & 078 & 1.66 & -0.96 & 16 & 12 & 23 \\
 2953.774 & 060 & 1.04 & -1.51 & 17 &  8 & 26 \\
 2986.617 & 254 & 3.41 & -2.07 & 30 & 41 & 22 \\
 2997.298 & 335 & 4.48 & -0.70 & 32 & 48 & 24 \\
          &     &      &       &    &    &    \\
 \hline
 \end{tabular}}
 \begin{centering}
 $^{*}$ If the absorption lines is blended or positioned at the end of an
 echelle no reliable determination of the profile parameters could be made
 and the item is left blank
 \end{centering}
 \end{table*}

 \section{Complete line identification of the available wavelength intervals
          from the CAT/CES spectra}
 \label{art2ap-optical}

 This appendix contains the figures of the optical CAT/CES spectra
 and a complete identification of several of these spectra. The spectra
 are ordered of increasing wavelength. As not all spectra are from
 the same date this means that one should be careful in comparing
 the spectra. From Sect.~\ref{art2sec-I}
 we know that there is some time variability
 of the observed radial velocities.
 All radial velocities quoted in this article are
 corrected to heliocentric velocities.

\begin{figure*}
\centerline{\hbox{\psfig{figure=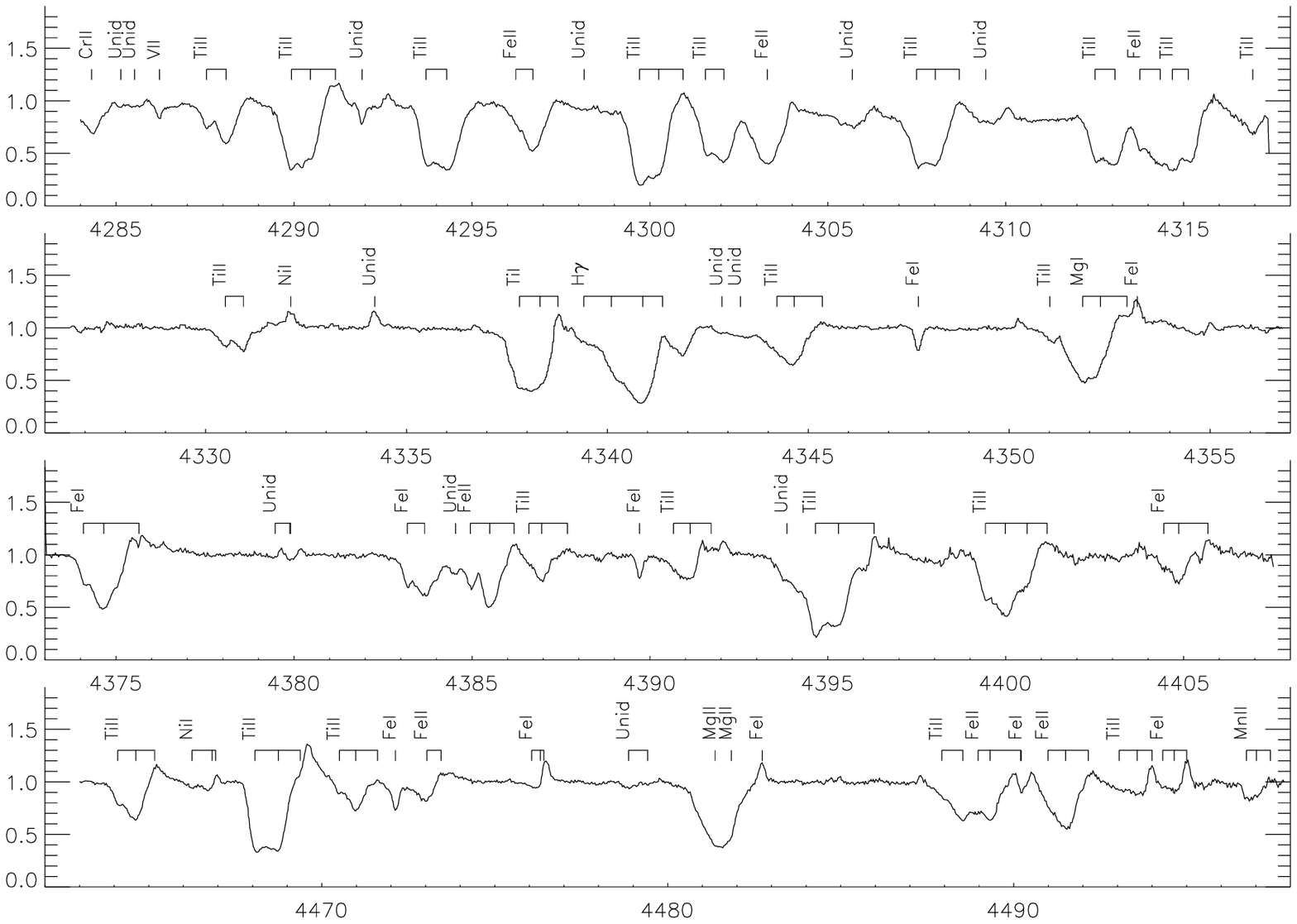,height=53pc,angle=90}}}
\caption{CAT/CES  spectrum of HD~101584}
\label{art2fig-opt1}
\end{figure*}

\begin{figure*}
\centerline{\hbox{\psfig{figure=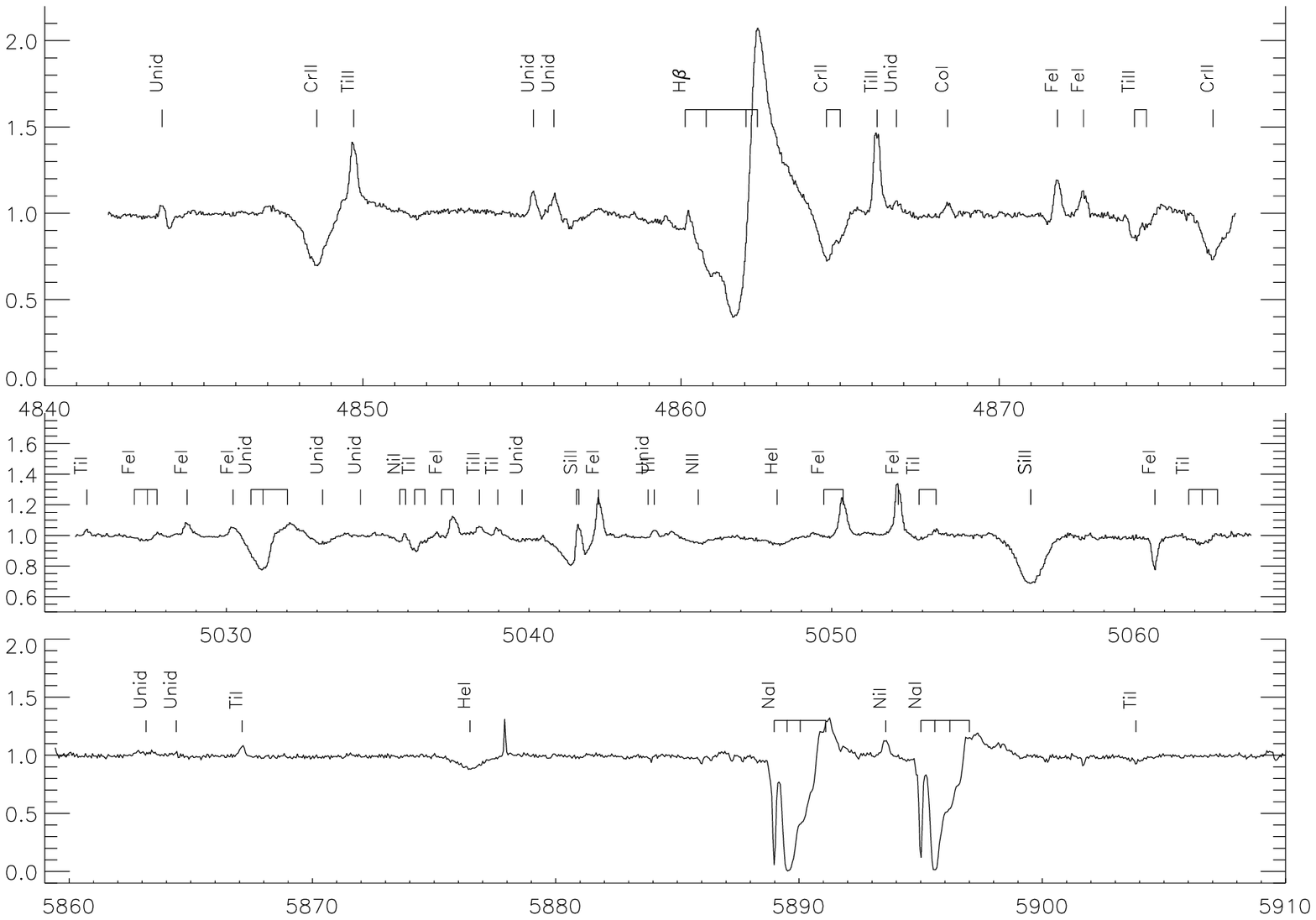,height=53pc,angle=90}}}
\caption{CAT/CES  spectrum of HD~101584}
\label{art2fig-opt2}
\end{figure*}

\begin{figure*}
\centerline{\hbox{\psfig{figure=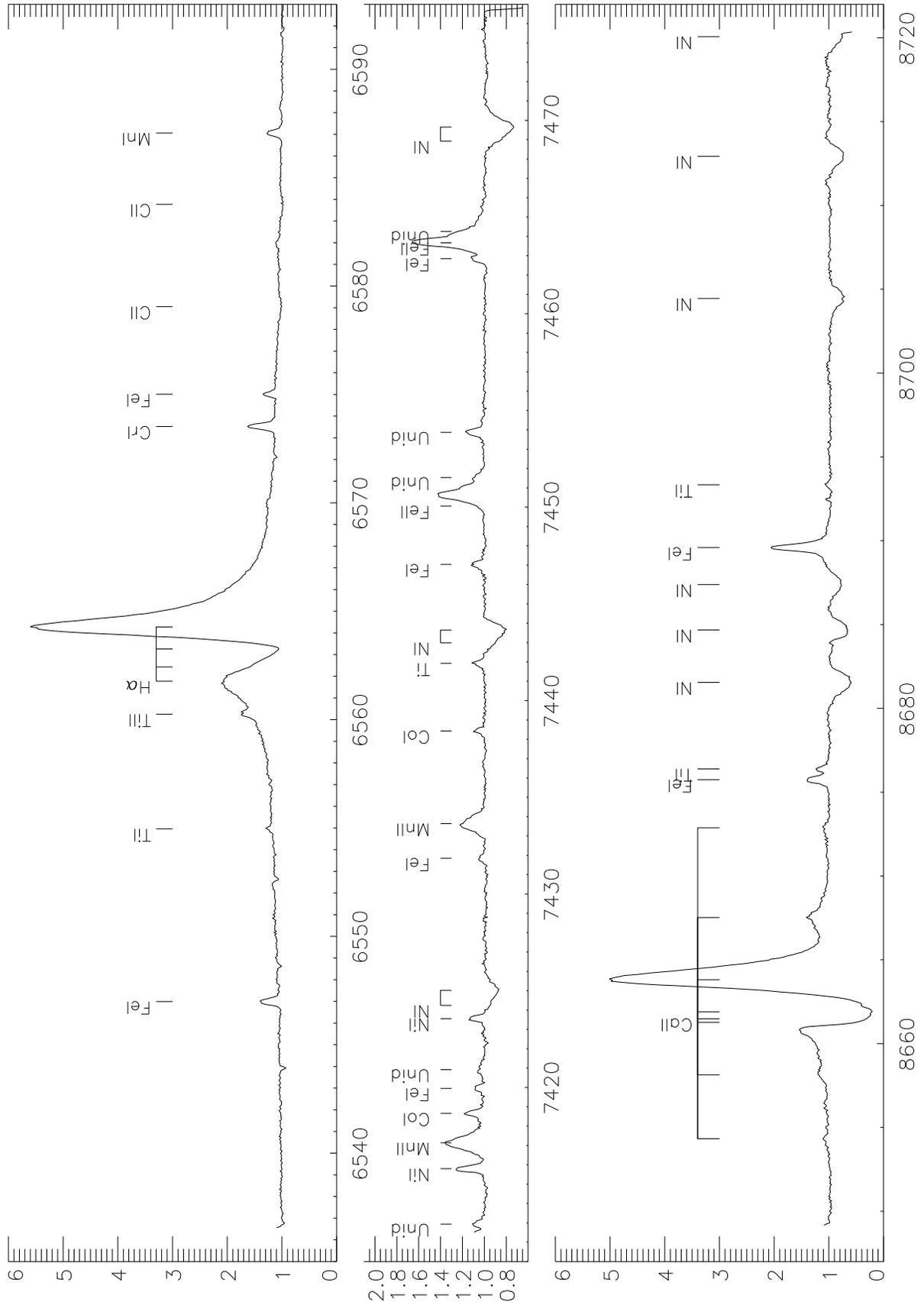,height=53pc,angle=90}}}
\caption{CAT/CES  spectrum of HD~101584}
\label{art2fig-opt3}
\end{figure*}

\begin{table*}
\caption{The optical spectrum of HD~101584}
\label{art2tab-optical}
\addtocounter{table}{-1}
\begin{tiny}
\centerline{\begin{tabular}{|llllrrrrrrrrrl|}
\hline
&                 &             &       &
&  I    &\multicolumn{3}{c}{II\&III\&IV}& V & VI &
&   & \\
\cline{7-9}
Ident.         & $\cs \lambda_{lab}$ & $\cs \chi$& $\log gf$&
$\cs \lambda_{obs}$ & abs. &abs. &abs & em & abs & em & $W$
&Depth & Remark \\
Ion(mtpl)      & [\AA]          & [eV]        &       &
        [\AA]          & \multicolumn{6}{c}{[km~s$^{-1}$]}
& [m\AA]& [\%]    &        \\
\hline
\multicolumn{14}{|c|}{4284-4318~\AA~ on January 20 1990}                     \\
\hline
CrII(31) & 4284.21 & 3.84-6.72 &-2.25 &4284.31&  & &26&  &  &  & 200&  30& \\
Unident  &         &           &      &4285.13&  & &  &  &  &  &  18&   6& \\
Unident  &         &           &      &4285.52&  & &  &  &  &  &  20&   6& \\
VII(23)  & 4286.13 & 1.65-4.56 &-2.81 &4286.22&  & &25&  &  &  &  43&  15&
blended \\
TiII(20) & 4287.893& 1.08-3.95 &-1.90 &4287.54&  &-6& &  &  &  &  85&  24& \\
         &         &           &      &4288.09&  & &33&  &  &  & 224&  40& \\
TiII(41) & 4290.222& 1.16-4.04 &-0.97 &4289.93&  &-1& &  &  &  & 380&  63& \\
         &         &           &      &4290.46&  & &36&  &  &  & 251&  49& \\
         &         &           &      &4291.16&  & &  &85&  &  & -23& -26& \\
Unident  &         &           &      &4291.91&  & &  &  &  &  &  49&  21& \\
TiII(20) & 4294.101& 1.08-3.95 &-1.05 &4293.71&  &-8& &  &  &  & 191&  43& \\
         &         &           &      &4294.29&  & &32&  &  &  & 560&  64& \\
FeII(28) & 4296.567& 2.69-5.57 &-3.36 &4296.23&  &-4& &  &  &  &    &  29& \\
         &         &           &      &4296.71&  & &29&  &  &  & 346&  47&  \\
Unident  &         &           &      &4298.15&  & &  &  &  &  &  23&   8&
noisy\\
TiII(41) & 4300.052& 1.18-4.05 &-0.47 &4299.71&  &-5& &  &  &  & 491&  77& \\
         &         &           &      &4300.25&  & &32&  &  &  & 341&  63& \\
         &         &           &      &4300.93&  & &  &80&  &  & -15&  -9& \\
TiII(41) & 4301.928& 1.16-4.02 &-1.28 &4301.56&  &-7& &  &  &  & 169&  37& \\
         &         &           &      &4302.08&  & &30&  &  &  & 412&  57& \\
FeII(27) & 4303.166& 2.69-5.56 &-2.53 &4303.30&  & &28&  &  &  & 557&  62& \\
Unident  &         &           &      &4305.69&  & &  &  &  &  &  91&  24&
broad \\
TiII(41) & 4307.900& 1.16-4.02 &-1.12 &4307.49&  &-9& &  &  &  & 234&  44& \\
         &         &           &      &4308.02&  & &27&  &  &  & 459&  58& \\
         &         &           &      &4308.69&  & &  &74&  &  & -34&  -6& \\
Unident  &         &           &      &4309.44&  & &  &  &  &  & 134&  21& \\
TiII(41) & 4312.861& 1.18-4.04 &-1.15 &4312.51&  &-5& &  &  &  & 290&  49& \\
         &         &           &      &4313.07&  & &33&  &  &  & 375&  56& \\
FeII(32) & 4314.289& 1.18-4.04 &-4.73 &4313.77&  & &(-17)&& &  &    &  47&
blended \\
         &         &           &      &4314.34&  & &23&  &  &  &    &  60&
blended \\
TiII(41) & 4314.979& 1.16-4.02 &-1.18 &4314.68&  &-2& &  &  &  &    &  66&
blended \\
         &         &           &      &4315.13&  & &30&  &  &  &    &  57&
blended \\
TiII(94) & 4316.807& 2.04-4.90 &-1.70 &4316.94&  & &28&  &  &  & 210&  30&  \\
         &         &           &      &       &  & &  &  &  &  &    &    & \\
\hline
\multicolumn{14}{|c|}{4327-4357~\AA~ on February 20 1989}                    \\
\hline
TiII(41) &4330.708 & 1.18-4.02 & -2.08&4330.49&  &1&  &  &  &  &  82&  17& \\
         &         &           &      &4330.94&  & &31&  &  &  &  70&  20& \\
NiI(52)  &4331.645 & 1.67-4.52 & -1.50&4332.12&  & &  &  &  &48& -43& -14& \\
Unident  &         &           &      &4334.21&  & &  &  &  &  & -40& -16&
LaII(24)?\\
TiI(20)  &4337.916 & 1.08-3.92 & -1.04&4337.81&  &8&  &  &  &  & 320&  53& \\
         &         &           &      &4338.32&  & &44&  &  &  & 294&  50& \\
         &         &           &      &4338.77&  & &  &75&  &  & -47& -21& \\
H$\cs\gamma$&4340.468&10.15-13.00&    &4339.41&  &(-58)&&&  &  &  44&  10& \\
         &         &           &      &4340.10&  &-10&&  &  &  &  44&  11& \\
         &         &           &      &4340.88&  & &44&  &  &  &1104&  71& \\
         &         &           &      &4341.37&  & &  &78&  &  &-189& -44& \\
Unident  &         &           &      &4342.85&  & &  &  &  &  &  14&   4& \\
Unident  &         &           &      &4343.31&  & &  &  &  &  &  47&   8& \\
TiII(20) &4344.291 & 1.08-3.92 & -2.00&4344.22&  &10& &  &  &  & 160&  18& \\
         &         &           &      &4344.65&  & &40&  &  &  & 146&  26& \\
         &         &           &      &4345.35&  & &  &89&  &  & -21&  -6& \\
FeI(2)   &4347.239 & 0.00-2.84 & -5.69&4347.74&  & &  &  &50&  &  42&  22& \\
TiII(94) &4350.834 & 2.05-4.89 & -3.11&4351.01&  & &(28)&&  &  &  26&  10&
blended \\
MgI(14)  &4351.906 & 4.33-7.16 & -1.72&4351.83&  &10& &  &  &  & 417&  47& \\
         &         &           &      &4352.27&  & &41&  &  &  & 119&  19& \\
         &         &           &      &4352.93&  & &  &86&  &  &-154& -15& \\
FeI(71)  &4352.737 & 2.21-5.05 & -1.32&4353.19&  & &  &  &  &47& -37& -15& \\
\hline
\multicolumn{14}{|c|}{4373-4407~\AA~ on January 21 1990}                     \\
\hline
FeI(648) &4374.495 & 3.29-6.11 & -2.54&4374.08&  &-9& &  &  &  &  50&  17& \\
         &         &           &      &4374.65&  & &30&  &  &  & 413&  52& \\
         &         &           &      &4375.64&  & &  &98&  &  & -40& -22& \\
Unident  &         &           &      &4379.47&  & &  &  &  &  &  18&   7&
NeII(56)? \\
         &         &           &      &4379.90&  & &  &  &  &  & 134&  33& \\
         &         &           &      &4379.88&  & &  &  &  &  &-157& -28& \\
FeI(41)  &4383.547 & 1.48-4.29 & -0.09&4383.18&  &-6& &  &  &  &  76&  22& \\
         &         &           &      &4383.67&  & &27&  &  &  & 255&  38& \\
Unident  &         &           &      &4384.54&  & &  &  &  &  &  71&  18& \\
FeII(27) &4385.381 & 2.77-5.58 & -2.63&4384.95&  &-10&&  &  &  &  88&  29& \\
         &         &           &      &4385.50&  & &27&  &  &  & 275&  50& \\
         &         &           &      &4386.18&  & &  &74&  &  & -31& -11& \\
TiII(104)&4386.858 & 2.59-5.40 & -0.79&4386.60&  &1&  &  &  &  &  14&   5& \\
         &         &           &      &4386.96&  & &26&  &  &  & 114&  24& \\
         &         &           &      &4387.68&  & &  &75&  &  & -21&  -7& \\
FeI(2)   &4389.244 & 0.05-2.86 & -4.57&4389.71&  & &  &  &51&  &  45&  22&
narrow \\
TiII(61) &4390.977 & 1.23-4.04 & -1.23&4390.66&  &-3& &  &  &  &  56&  10& \\
         &         &           &      &4391.13&  & &30&  &  &  & 136&  23& \\
         &         &           &      &4391.72&  & &  &70&  &  &-181& -25& \\
Unident  &         &           &      &4393.85&  & &  &  &  &  & 135&  24& \\
TiII(19) &4395.031 & 1.06-3.89 & -0.55&4394.65&  &-7& &  &  &  & 476&  71& \\
         &         &           &      &4395.30&  & &37&  &  &  & 449&  62& \\
         &         &           &      &4396.30&  & &  &106& &  & -82& -19& \\
TiII(51) &4399.767 & 1.23-4.04 & -1.32&4399.43&  &-4& &  &  &  & 146&  33& \\
         &         &           &      &4399.99&  & &34&  &  &  & 420&  58& \\
         &         &           &      &4400.60&  & &(76)&&  &  &  99&  25& \\
         &         &           &      &4401.16&  & &  &114& &  & -52& -12& \\
FeI(41)  &4404.752 & 1.55-4.35 & -0.57&4404.44&  &-2& &  &  &  &  41&  13& \\
         &         &           &      &4404.86&  & &26&  &  &  & 119&  26& \\
         &         &           &      &4405.68&  & &  &82&  &  & -73& -13& \\
\hline
\multicolumn{14}{|c|}{4463-4498~\AA~ on January 20 1990}                     \\
\hline
TiII(40) &4464.458 & 1.16-3.92 & -1.62&4464.10&  &-5& &  &  &  & 40 &  14& \\
         &         &           &      &4464.63&  & &30&  &  &  &263 &  36& \\
         &         &           &      &4465.17&  & &  &67&  &  &-118& -20& \\
NiI(168) &4466.394 & 3.69-6.45 & +0.02&4466.25&  &9&  &  &  &  & 20 &   6& \\
         &         &           &      &4466.83&  & &48&  &  &  & 59 &  13& \\
         &         &           &      &4466.93&  & &  &55&  &  &-46 & -17& \\
TiII(31) &4468.493 & 1.13-3.89 & -0.63&4468.07&  &-9& &  &  &  &221 &  53& \\
         &         &           &      &4468.75&  & &36&  &  &  &656 &  74& \\
         &         &           &      &4469.38&  & &  &79&  &  &-425& -48& \\
TiII(40) &4470.864 & 1.16-3.95 & -2.05&4470.51&  &-5& &  &  &  & 21 &   9& \\
         &         &           &      &4470.98&  & &27&  &  &  &138 &  27& \\
         &         &           &      &4471.61&  & &  &69&  &  & -4 &  -2& \\
FeI(2)   &4471.68  & 0.11-2.87 & -6.20&4472.13&  & &  &  &49&  & 69 &  26&
narrow \\
\hline
\end{tabular}}
\end{tiny}
\end{table*}

\begin{table*}
\caption{continued}
\addtocounter{table}{-1}
\begin{tiny}
\centerline{\begin{tabular}{|llllrrrrrrrrrl|}
\hline
&                 &             &       &
&  I    &\multicolumn{3}{c}{II\&III\&IV}& V & VI &
&   & \\
\cline{7-9}
Ident.         & $\cs \lambda_{lab}$ & $\cs \chi$&$\log gf$&
$\cs \lambda_{obs}$ & abs. &abs. &abs & em & abs & em & $W$
&Depth & Remark \\
Ion(mtpl)      & [\AA]          & [eV]        &       &
        [\AA]          & \multicolumn{6}{c}{[km~s$^{-1}$]}
& [m\AA]& [\%]    &        \\
\hline
FeII(37) &4472.921 & 2.83-5.59 & -4.31&4473.04&  & &27&  &  &  &148 &  25& \\
         &         &           &      &4473.45&  & &  &55&  &  &-115& -13& \\
FeI(350) &4476.021 & 2.83-5.59 & -0.67&4476.07&  & &22&  &  &  & 18 &   4& \\
         &         &           &      &4476.32&  & &39&  &  &  & 46 &  18& \\
         &         &           &      &4476.42&  & &  &46&  &  &-91 & -28& \\
Unident  &         &           &      &4478.87&  & &  &  &  &  & 25 &   6& \\
         &         &           &      &4479.42&  & &  &  &  &  &  7 &   2& \\
MgII(4)  &4481.129 & 8.83-11.58& +0.74&4481.37&  & &35&  &  &  &614 &  59& \\
MgII(4)  &4481.327 & 8.83-11.58& +0.59&4481.84&  & &53&  &  &  & 34 &  19& \\
FeI(68)  &4482.257 & 2.21-4.97 & -1.45&4482.73&  & &  &  &  &51&-45 & -21& \\
TiII(115)&4488.319 & 3.11-5.86 & -0.62&4487.92&  &-8& &  &  &  & 62 &  11&
blended \\
         &         &           &      &4488.53&  & &33&  &  &  &241 &  36&
blended \\
FeII(37) &4489.185 & 2.82-5.57 & -3.50&4488.97&  &5&  &  &  &  & 44 &  13&
blended \\
         &         &           &      &4489.32&  & &28&  &  &  &166 &  35&
blended \\
         &         &           &      &4490.20&  & &  &87&  &  &-118& -29& \\
FeI(2)   &4489.741 &  0.12-2.86& -3.98&4490.22&  & &  &  &51&  &108 &  38&
narrow \\
FeII(37) &4491.401 &  2.84-5.59& -2.89&4490.99&  &-8& &  &  &  & 45 &  13& \\
         &         &           &      &4491.50&  & &26&  &  &  &341 &  43& \\
         &         &           &      &4492.16&  & &  &70&  &  &-53 & -12& \\
TiII(18) &4493.53  &  1.08-3.82& -3.62&4493.05&  &-13&&  &  &  & 38 &   8& \\
         &         &           &      &4493.57&  & &22&  &  &  & 55 &  12& \\
         &         &           &      &4494.00&  & &  &50&  &  &-29 & -18& \\
FeI(68)  &4494.568 &  2.19-4.93& -1.07&4494.31&  &2&  &  &  &  & 10 &   5& \\
         &         &           &      &4494.64&  & &24&  &  &  & 30 &  10& \\
         &         &           &      &4495.00&  & &  &48&  &  &-45 & -23& \\
MnII(17) &4496.989 &10.60-13.35& -0.92&4496.73&  &2&  &  &  &  & 23 &  10& \\
         &         &           &      &4497.02&  & &21&  &  &  & 72 &  13& \\
         &         &           &      &4497.42&  & &  &48&  &  &-23 &  -8& \\
\hline
\multicolumn{14}{|c|}{4842-4877~\AA~ on  February 20 1989}                   \\
\hline
Unident  &         &           &      &4843.69&  & &  &  &  &  & -7 &  -5& \\
CrII(30) &4848.24  & 3.85-6.39 & -1.40&4848.55&  & &35&  &  &  &222 &  30& \\
TiII(29) &4849.18  & 1.13-3.67 & -2.98&4849.71&  & &  &  &  &48&-93 & -30& \\
Unident  &         &           &      &4855.36&  & &  &  &  &  &-26 & -13&
YII(22)? \\
Unident  &         &           &      &4856.00&  & &  &  &  &  &-24 & -10& \\
H$\cs\beta$&4861.332&10.15-12.69&     &4860.13&  &(-58)&&&  &  & 80 &  11& weak
\\
         &         &           &      &4860.79&  &-18&&  &  &  &181 &  27& \\
         &         &           &      &4862.04&  & &(60)&&  &  &1163&  88& \\
         &         &           &      &4862.40&  & &  &82&  &  &-1266&-175& \\
CrII(30) &4864.32  & 3.84-6.39 & -1.66&4864.57&  & &31&  &  &  &130 &  27& \\
         &         &           &      &4865.00&  & &(58)&&  &  & 45 &  12& \\
TiII(29) &4865.620 & 1.11-3.65 & -2.72&4866.16&  & &  &  &  &49&-124& -49& \\
Unident  &         &           &      &4866.77&  & &  &  &  &  & -22&  -6& \\
CoI(158) &4867.870 & 3.10-5.65 & 0.27 &4868.38&  & &  &  &  &47& -12&  -6& \\
FeI(318) &4871.323 & 2.85-5.39 &-0.54 &4871.83&  & &  &  &  &47& -48& -20& \\
FeI(318) &4872.144 & 2.87-5.40 &-0.28 &4872.65&  & &  &  &  &46& -32& -12& \\
TiII(114)&4874.025 & 3.08-5.88 &-0.99 &4874.25&  & &30&  &  &  &  48&  14& \\
         &         &           &      &4874.63&  & &(53)&&  &  &  26&   8& \\
CrII(30) &4876.41  & 3.84-6.37 &-1.70 &4876.72&  & &34&  &  &  & 178&  26& \\
\hline
\multicolumn{14}{|c|}{5025-5064~\AA~ on February 19 1989}                    \\
\hline
TiI(38)  &5024.842 & 0.81-3.29 & -0.52&5025.39&  & &  &  &  &48&  -7&  -4& \\
FeI(1065)&5027.136 & 4.14-6.59 & -0.91&5026.96&  &5&  &  &  &  &   8&   2& \\
         &         &           &      &5027.39&  & &31&  &  &  &   1&   3& \\
         &         &           &      &5027.71&  & &  &50&  &  &  -8&  -3& \\
FeI(791) &5028.129 & 3.56-6.01 & -1.34&5028.70&  & &  &  &  &50& -23&  -8& \\
FeI(718) &5029.623 & 3.40-5.85 & -1.90&5030.22&  & &  &  &  &51& -15&  -5& \\
Unident  &         &           &      &5030.81&  & &  &  &  &  &  34&   8& \\
         &         &           &      &5031.21&  & &  &  &  &  & 118&  22& \\
         &         &           &      &5032.02&  & &  &  &  &  & -67&  -1& \\
Unident  &         &           &      &5033.18&  & &  &  &  &  &  35&   6&
CII(17)? \\
Unident  &         &           &      &5034.43&  & &  &  &  &  &  24&   3& hot
star \\
NiI(143) &5035.374 & 3.62-6.07 & +0.38&5035.73&  & &37&  &  &  &  15&   4& \\
         &         &           &      &5035.92&  & &  &48&  &  & -12&  -5& \\
TiI(110) &5036.468 & 1.44-3.89 & +0.32&5036.22&  &1&  &  &  &  &  42&   1& \\
         &         &           &      &5036.56&  & &21&  &  &  &  11&   3& \\
FeI(465) &5036.931 & 3.00-5.45 & -3.30&5037.11&  & &26&  &  &  &   4&   2& \\
         &         &           &      &5037.50&  & &  &50&  &  & -36& -13& \\
TiII(71) &5037.81  & 1.57-4.02 & -3.56&5038.36&  & &  &  &  &48& -17&  -6& \\
TiI(110) &5038.400 & 1.42-3.87 & +0.21&5038.97&  & &  &  &  &50& -11&  -5& \\
Unident  &         &           &      &5039.77&  & &  &  &  &  &  29&   5& hot
star \\
SiII(5)  &5041.063 &10.02-12.47& +0.28&5041.57&  & &46&  &  &  & 251&  20& \\
         &         &           &      &5041.65&  & &  &51&  &  & -62& -30& \\
FeI(36)  &5041.759 & 1.48-3.93 & -2.78&5042.30&  & &  &  &  &48& -81& -33& \\
Unident  &         &           &      &5043.94&  & &  &  &  &  &  17&   2& hot
star \\
TiI(38)  &5043.578 & 0.83-3.28 & -1.65&5044.14&  & &  &  &  &49& -13&  -5& \\
NII(4)   &5045.098 &18.40-20.85& -0.39&5045.59&45& & &  &  &  &  44&    5& \\
HeI(47)  &5047.736 &21.13-23.57& -1.60&5048.20&43& & &  &  &  &  59&    6& \\
FeI(114) &5049.825 & 2.27-4.71 & -1.32&5049.74&  & &11&  &  &  &  11&   2& \\
         &         &           &      &5050.38&  & &  &49&  &  & -68& -23& \\
FeI(16)  &5051.636 & 0.91-3.35 & -2.45&5052.20&  & &  &  &  &49& -85& -33& \\
TiI(199) &5052.879 & 2.17-4.61 & -0.31&5052.89&  & &16&  &  &  &  11&   3& \\
         &         &           &      &5053.45&  & &  &50&  &  & -11&  -4& \\
SiII(5)  &5056.020 &10.03-12.47& +0.54&5056.58&  & &  &49&  &  & 332&  32&
broad\\
SiII(5)  &5056.353 &10.03-12.47& -0.42&5056.58&  & &  &  &  &  &    &    & \\
FeI(1)   &5060.079 & 0.00-2.44 & -6.16&5060.68&  & &  &  &51&  &  49&  21&
narrow \\
TiI(199) &5062.112 & 2.15-4.59 & -0.47&5061.80&  &-3& &  &  &  &   3&   1& \\
         &         &           &      &5062.24&  & &23&  &  &  &  34&   5& \\
         &         &           &      &5062.75&  & &  &53&  &  &  -8&  -3& \\
\hline
\multicolumn{14}{|c|}{5859-5911~\AA on February 13 1993}                    \\
\hline
Unident  &         &           &      &5863.16&  & &  &  &  &  & -40&  -5&
broad \\
Unident  &         &           &      &5864.40&  & &  &  &  &  & -12&  -3& \\
TiI(72)  &5866.453 &10.6-3.17  &-0.80 &5867.11&  & &  &  &  &51& -19&  -8& \\
HeI(11)  &5875.618 &20.87-22.97&+0.74 &5876.48&61& &  &  &  &  & 135&  12& \\
NaI(1)D2 &5889.953 & 0.00-2.10 &+0.11 &5888.98&  & &  &  &(-33)&&152&  83& CS
\\
         &         &           &      &5889.51&  &-6& &  &  &  & 405&  75& \\
         &         &           &      &5890.05&  & &22&  &  &  & 583&  57& \\
         &         &           &      &5891.10&  & &  &75&  &  &-271& -33& \\
NiI(68)  &5892.878 & 1.98-4.07 &-1.36 &5893.57&  & &  &  &  &52& -39& -14& \\
NaI(1)D1 &5895.923 & 0.00-2.09 &-0.19 &5895.01&  & &  &  &(-30)&&160&  88& CS
\\
         &         &           &      &5895.58&  &-1& &  &  &  & 409&  80& \\
         &         &           &      &5896.20&  & &31&  &  &  & 512&  46& \\
         &         &           &      &5897.00&  & &  &72&  &  &-277& -27& \\
TiI(71)  &5903.317 & 1.06-3.15 &-1.90 &5903.85&  & &  &  &(44)&&  41&   5&
wrong? \\
\hline
\end{tabular}}
\end{tiny}
\end{table*}

\begin{table*}
\caption{continued}
\addtocounter{table}{-1}
\begin{tiny}
\centerline{\begin{tabular}{|llllrrrrrrrrrl|}
\hline
&                 &             &       &
&  I    &\multicolumn{3}{c}{II\&III\&IV}& V & VI &
&   & \\
\cline{7-9}
Ident.         & $\cs \lambda_{lab}$ & $\cs \chi$&$\log gf$&
$\cs \lambda_{obs}$ & abs. &abs. &abs & em & abs & em & $W$
&Depth & Remark \\
Ion(mtpl)      & [\AA]          & [eV]        &       &
        [\AA]          & \multicolumn{6}{c}{[km~s$^{-1}$]}
& [m\AA]& [\%]    &        \\
\hline
\hline
\multicolumn{14}{|c|}{6537-6593~\AA~ February 12 1993}                       \\
\hline
FeI(268) &6546.245 & 2.75-4.63 &-1.72 &6546.99&  & &  &  &  &52& -21& -4& \\
TiI(102) &6554.226 & 1.44-3.32 &-1.28 &6554.96&  & &  &  &  &51& -22& -11& \\
TiII(91) &6559.580 & 2.04-3.92 &-2.14 &6560.25&  & &  &  &  &48& -54& -20& \\
H$\cs\alpha$&6562.817&10.15-12.04&    &6561.77&  &-31&&  &  &  & &       &
edge\\
         &         &           &      &6562.43&  &-1& &  &  &  &    &    & \\
         &         &           &      &6563.26&  & &37&  &  &  &    &    & abs.
min.  \\
         &         &           &      &6564.27&  & &  &83&  &  &    &    & abs.
max.  \\
CrI(16)  &6572.900 & 1.00-2.88 &-4.01 &6573.52&  & &  &  &  &45&-165& -39& \\
FeI(13)  &6574.238 & 0.99-2.85 &-5.02 &6575.01&  & &  &  &  &52& -54& -20& \\
CII(2)   &6578.03  &14.39-16.26&      &6579.05&64& &  &  &  &  &  85&   5& \\
CII(2)   &6582.85  &14.39-16.26&      &6583.77&59& &  &  &  &  &  45&   2& \\
MnI(51)  &6586.343 & 4.41-6.28 &-2.01 &6587.06&  & &  &  &  &50& -91& -29& \\
\hline
\multicolumn{14}{|c|}{7413-7476~\AA~ on April 20 1992}                       \\
\hline
Unident  &         &           &      &7412.93&  & &  &  &  &  & -41& -11& \\
NiI(62)  &7414.51  &1.98-3.64  &-2.77 &7415.80&  & &  &  &  &51& -95& -28& \\
MnII(4)  &7415.78  &3.69-5.35  &-2.46 &7417.14&  & &  &  &  &54&-306& -36&
broad \\
CoI(89)  &7417.38  &2.03-3.70  &-2.42 &7418.66&  & &  &  &  &51& -69& -19& \\
FeI(1001)&7418.674 &4.12-5.79 &-1.17  &7419.95&  & &  &  &  &50& -38& -10& \\
Unident  &         &           &      &7420.92&  & &  &  &  &  & -44&  -6& \\
NiI(139) &7422.30  &3.62-5.28  &+0.06 &7423.56&  & &  &  &  &50& -46& -15& \\
NI(3)    &7424.24  &10.28-11.94&-0.61 &7424.24&(23)&& &  &  &  &  13&   4& \\
         &         &           &      &7425.02&55& &  &  &  &  & 123&  12& \\
FeI(204) &7430.58  &2.58-4.24  &-3.81 &7431.85&  & &  &  &  &50& -21&  -6& \\
MnII(4)  &7432.27  &3.69-5.35  &-2.79 &7433.64&  & &  &  &  &54&-199& -23&
broad \\
CoI(53)  &7437.16  &1.95-3.61  &-3.64 &7438.43&  & &  &  &  &50& -34& -10& \\
Ti(225)  &7440.600 &2.25-3.90  &-1.19 &7441.93&  & &  &  &  &52& -34& -10& \\
NI(3)    &7442.28  &10.29-11.94&-0.31 &7442.99&(27)&& &  &  &  &  58&   8& \\
         &         &           &      &7443.64&54& &  &  &  &  & 142&  19& \\
FeI(1077)&7445.776 &4.24-5.90 &-0.13  &7447.04&  & &  &  &  &50& -52& -13& \\
FeII(73) &7449.34  &3.87-5.53  &-3.60 &7450.06&  & &  &  &  &51&-332& -42&
broad \\
Unident  &         &           &      &7451.53&  & &  &  &  &  & -50& -10&
YII(25)? \\
Unident  &         &           &      &7453.87&  & &  &  &  &  & -70& -18& \\
FeI(204) &7461.534 &2.55-4.20  &-3.48 &7462.85&  & &  &  &  &52& -44& -13& \\
FeII(73) &7462.38  &3.87-5.53  &-2.98 &7463.67&  & &  &  &  &51&-377& -63&
broad \\
Unident  &         &           &      &7464.25&  & &  &  &  &  &-159& -20& \\
NI(3)    &7468.29  &10.28-11.94&-0.13 &7468.92&(24)&& &  &  &  &  63&   8& \\
         &         &           &      &7469.65&53& &  &  &  &  & 251&  25& \\
\hline
\multicolumn{14}{|c|}{8649-8721~\AA~ on February 13 1993}                    \\
\hline
CaII(2)  &8662.140 &~1.69-3.11 &-0.73 &8661.48&  &-6& &  &  &  &    &  71& rel.
min.\\
         &         &           &      &8661.90&  & &9 &  &  &  &    &  77& abs.
min.\\
         &         &           &      &8661.27&  & &21&  &  &  &    &  65& rel.
min.\\
         &         &           &      &8663.81&  & &  &75&  &  &    &-400& abs.
max.\\
         &         &           &      &8658.14&  & &  &  &  &(-122)&& -18&-172
km/s\\
         &         &           &      &8667.52&  & &  &  &  &( 203)&& -37& 153
km/s \\
         &         &           &      &8654.33&  & &  &  &  &(-253)&&  -1&-303
km/s \\
         &         &           &      &8672.87&  & &  &  &  &( 339)&&  -6& 289
km/s \\
FeI(339) &8674.751 & 2.82-2.42 &-1.89 &8675.74&  & &  &  &  &51&-183& -41& \\
TiI(68)  &8675.38  & 1.06-2.48 &-1.30 &8676.39&  & &  &  &  &52& -87& -24& \\
NI(1)    &8680.24  &10.29-11.71&+0.42 &8681.55&62& &  &  &  &  & 525&  40& \\
NI(1)    &8683.38  &10.29-11.71&+0.14 &8684.68&62& &  &  &  &  & 435&  35& \\
NI(1)    &8686.13  &10.29-11.70&-0.27 &8687.38&60& &  &  &  &  & 299&  22& \\
FeI(60)  &8688.633 & 2.17-3.59 &-1.41 &8689.59&  & &  &  &  &50&-550&-107& \\
TiI(68)  &8692.34  & 1.04-2.46 &-1.92 &8693.34&  & &  &  &  &52& -15&  -6& \\
NI(1)    &8703.24  &10.29-11.70&-0.27 &8704.45&59& &  &  &  &  & 365&  27& \\
NI(1)    &8711.69  &10.29-11.70&-0.16 &8712.93&60& &  &  &  &  & 329&  28& em.
on wings \\
NI(1)    &8718.82  &10.29-11.71&-0.23 &8720.07&60& &  &  &  &  & 280&  23& \\
\hline
\end{tabular}}
\end{tiny}
\end{table*}

\hfill

\end{document}